%% file: NP-MScThesis.tex
\documentclass[aps,twocolumn,showpacs,showkeys,amsmath,amssymb,superscriptaddress,floatfix]{revtex4}
\usepackage{graphicx}% Include figure files
\usepackage{dcolumn}% Align table columns on decimal point
\usepackage{bm}
\begin{document}
\title{Baryogenesis at the electroweak phase transition}
\author{Nicholas Petropoulos}
\email{nicholas@teor.fis.uc.pt} 
\affiliation{Department of Physics and
Astronomy, University of Manchester, Manchester M13 9PL, United Kingdom} 
\affiliation{Centro de F\'{\i}sica Te\'{o}rica, Departamento de
F\'{\i}sica, Universidade de Coimbra, P3004--516 Coimbra, Portugal}
\date{29 April 2003\footnote{\input{note}}}

\begin{abstract}
A possible solution to the observed baryon asymmetry in the universe is 
described, based on the physics of the standard model of electroweak 
interactions. At temperatures high enough electroweak physics provides
violation of baryon number, while $C$ and $CP$ symmetries are not exactly
conserved, although in the context of the minimal electroweak model with one
Higgs doublet the rate of $CP$ violation is not sufficient enough to 
generate the observed asymmetry. The condition that the universe must be out 
of thermal equilibrium requires the electroweak phase transition (EWPT) to be first
order. The dynamics of the phase transition in the minimal model is 
investigated  through 
the effective potential, which is calculated at the one loop order. Finite 
temperature effects on the effective potential are treated 
numerically and within the 
high temperature approximation, which is found to be in good agreement with 
the exact calculation. At the one loop level the phase transition was found
to be of the first order, while the strength of the transition depends
on the unknown parameters of the theory which are the Higgs boson and top
quark masses \cite{top}~.
\end{abstract} 

%\preprint{UC-20-2003} 
%11.10.Wx  Finite-temperature field theory
%11.15.Ex  Spontaneous breaking of gauge symmetries
%11.30.Qc  Spontaneous and radiative symmetry breaking
%11.30.Fs  Global symmetries (e.g., baryon number, lepton number)  
%12.15.-y  Electroweak interactions
%12.15.Ji  Applications of electroweak models to specific processes 
%98.80.Ft  Origin, formation, and abundances of the elements (see also 26.35.+c Big Bang nucleosynthesis
%                 in nuclear astrophysics)
%98.80.Cq  Particle-theory and field-theory models of the early Universe (including cosmic pancakes,
%                 cosmic strings, chaotic phenomena, inflationary universe, etc.) (see also 11.25.-w Strings
%                 nd branes, and 11.10.-z in general theory of fields and particles) 
%         
%Use showkeys class option if keyword display desired
\maketitle
\tableofcontents

\noindent
\textit{``Why does the whole world have $<\phi>=+v$? Why doesn't it have 
$<\phi>=-v$ somewhere? Suppose that God created the universe in the
state  $<\phi>=0$  and then the universe discovered that it could lower 
its energy;  Where it puts its energy is none of my business, but it gets rid
of it -- gives it back to God  or something;''}

\hspace {5.7cm} R.~Feynman \cite{Feynman:1976}

\input{chap1}

\input{chap2}

\input{chap3}
\input{chap4}

\input{chap5}

\input{chap6}
\input{remarks}

\input{acknow}

\input{bib-spires}
\end{document}

%% file: note.tex
The initial report on which this work is based, has been submitted
in a form of a thesis, in order to fulfil the requirements of a MSc degree 
in theoretical physics, at the University of Manchester on April 1994. However, since there
has been quite some progress on electroweak baryogenesis during the last 
few years, this work might seem to be slightly
out of date and I have not 
attempted  to really revise it. I have only tried to add some references to 
recent reviews on the subject. However, since it contains some 
introductory material, I thought (and I hope) that it might be useful to someone
and so I decided to put this in the arXiv.  

%% file: chap1.tex
\section{\label{Sect:Baryogenesis-Problem}The Baryogenesis Problem}

\subsection{Overview}

Recently the connection between particle physics and cosmology has
received much attention. The classical cosmological model of the expanding universe
provides a powerful framework where particle physics can test its
predictions about matter genesis. The expansion of the universe can be 
considered as an enormous particle accelerator, which although ran billions
of years ago in the far past, can still be used to check the validity of
theories concerning the elementary particles. Cosmology on the other
hand can use the predictions of particle physics about the nature and 
behaviour of the elementary particles in order to cure unsolved up to now 
cosmological problems which are involved in the theories concerned with the 
evolution of the universe. 

The baryogenesis problem, that is the observed excess of
matter over antimatter today in the universe, is one of these problems. As 
pointed out by Sakharov 
in 1967 \cite{Sakharov:dj}, the universe began in a baryon symmetric 
state but particle interactions produced a net asymmetry. He postulated 
three conditions to be satisfied in order to explain the observed baryon excess. These 
are: (a) baryon number non--conservation, (b) $C$ and $CP$ violation
and (c) that the universe must be out of thermal equilibrium. Our aim is to 
explore how these conditions are satisfied in the
framework of the standard model of electroweak interactions. We will focus
our attention on the third condition and provide a careful study of the
electroweak phase transition. Our main goal is to examine the order of this
transition when temperature is introduced in the model via the so--called one 
loop approximation. 

This work is organized as follows: 
In order to establish the basic principles for the baryogenesis problem 
the remainder of this section is devoted to a brief account on the development
of the big bang model, the standard cosmological model, from its beginning
up to the present days. The most important stages in the development of the 
big bang model are given in Section~\ref{Sect:Big-Bang}. In Section~\ref{Sect:Baryon-Asymmetry} we 
present how the 
baryogenesis problem appears in the framework of the standard big bang 
cosmology. The necessity to satisfy the Sakharov conditions and relevant
comments are outlined in Section~\ref{Sect:3-Conditions}. A recent development in the
standard hot big bang model, the inflation model, is discussed briefly in 
Section~\ref{Sect:Inflation}.

According to recent investigations it is possible that the
necessary conditions for the solution of the baryogenesis problem can be
satisfied in the standard model of the electroweak interactions, so we find 
very useful to provide an introduction to
the standard model of these interactions and
its connection with cosmology. 
This is the subject of Section~\ref{Sect:Gauge-Theories}, where phase transitions in 
gauge theories are also discussed briefly
through well known examples. 

The standard electroweak model satisfies the first two conditions, so
we sketch in Section~\ref{Sect:BCP-Violation} how baryon number and $CP$ symmetry are 
violated in the model. In the standard  model, due to 
the non--trivial structure of the $SU(2)$ gauge vacuum and as a consequence 
of anomalous processes, baryon number is not conserved. On the other hand 
$CP$ symmetry is violated in the standard model because of relative phases 
between the electroweak gauge interactions and the Higgs interactions 
of the quarks. 

The third condition is satisfied if the
electroweak phase transition (EWPT) is of the first order. 
The basic tool for the investigation of the electroweak phase
transition is the finite temperature effective potential, the quantity which
has the meaning of the free energy density of the system under consideration,
and the relevant theory is given in some detail in Section~\ref{Sect:Veff}. 

The evolution and the order of the electroweak phase 
transition will be determined in the so--called one loop approximation 
in Section~\ref{Sect:EWPT}. Finite temperature effects are treated exactly and within the high
temperature limiting case. 

The last section is devoted  to a discussion concerning 
our investigation of the electroweak phase transition in connection
with the baryogenesis problem and we also present 
there a summary of our results and conclusions.     

\subsection{\label{Sect:Big-Bang}The Hot Big Bang}

Modern cosmological theories started to develop within the framework 
of Einstein's general theory of relativity early this century.  
The development of particle physics in the recent years has infused
new ideas in cosmology, which can be applied to the study of the
earliest moments of the universe and the evolution to its present state.
The most accepted up to date cosmological 
theory is the hot big bang model. This theory is based upon general relativity  
and the Friedmann model for an expanding universe. According to 
this model the universe starts its life from a state of 
enormous matter density and temperature and after its expansion results
in what we observe today. The evolution and the most important stages 
in the development of this 
cosmological theory up to the present days have been explored in 
an elegant way by many authors \cite{Hawking:1988,Weinberg:1976,Parker:1988}.

A detailed discussion on the formulation of this theory is given
by Weinberg \cite{Weinberg:1972} and Kolb and Turner \cite{Kolb:1990}, while a 
brief account can be found in          
Linde \cite{Linde:1990}. The beginning of the theory and the first 
important stage goes back in 1922  when 
Friedmann  based  on Einstein's cosmological principle, which is the
hypothesis that the universe is spatially homogeneous and isotropic, and 
created a model for an expanding universe. A few years later, Hubble's and 
Slipher's 
observations of the galactic redshift supported experimentally the idea of
the expansion from an initial state and very soon this theory became the 
cornerstone of modern cosmology.

In Friedmann's model the radius of the universe $R=R(t)$, or more 
explicitly its scale factor, is a function of time and has an evolution 
which is described by the  equations of general relativity. Einstein's 
equations imply  an evolution for the scale factor given by the first order
differential equation 
\begin{equation}
\dot{R}+k=\frac{8\pi G}{3}\rho R^{2}~.
\end{equation}
The constant $k$ in the above equation describes the curvature of space and 
takes the values $k=-1,\;0,\; 1$, which correspond to open, flat and 
close universe models
respectively and $G$ is the gravitational constant. In addition we have
an energy conservation law which can be expressed by
\begin{equation}
\dot{\rho}R^{3}+3(\rho+p)R^{2}\dot{R}=0~,
\end{equation}            
where $\rho$ is the energy density of matter in the universe and $p$ its 
pressure. In order to solve these equations and find out how the universe evolves in 
time, we need a state equation which relates the energy density
of the universe to its pressure. We may assume $p=\omega\rho$ so 
the above equations give the result
\begin{equation}
\rho \sim R^{-3(1+\omega)}~.
\end{equation}
We can consider two possibilities, in order to interpret this solution. If
we suppose
that the universe was filled with nonrelativistic cold matter with 
negligible pressure $p=0$, we find that
$\rho \sim R^{-3}$. If the universe was
a hot ultrarelativistic gas of noninteracting particles with $p=\rho/3$, we 
then find that $\rho \sim R^{-4}$. When $R$ is small one finds that for nonrelativistic cold 
matter the radius of the universe evolves in time as $R \sim t^{2/3}$, while 
in the case of the hot ultrarelativistic gas it goes as  
$R \sim t^{1/2}$.
Thus regardless the model used, as the time $t$ tends to zero, the scale 
factor vanishes and the density of matter becomes infinite. The point 
$t=0$ is  known as  the initial cosmological singularity. It is the initial 
point when the universe starts its life and then expands to what it is
today. The expansion rate of the universe is given by the quantity
$H=\dot{R}/R$ which is known as the Hubble constant, although it is not 
really a constant but it varies with time.

Up to the mid--sixties it was not clear if the universe had started its 
life from a hot or a cold initial state.
The new era in cosmology opened when Penzias and Wilson discovered the $2.7^{0}
K$ microwave background radiation  in 1964--65, which had been predicted by 
the hot universe theory.  If we suppose that the 
universe expanded adiabatically, then during the expansion the quantity $RT$
remained approximately constant, and the temperature of the universe dropped
off as  $T \sim R^{-1}(t)$.
The radiation which was  detected by Penzias and Wilson 
were the relic photons of the initially hot photon gas which  cooled 
down during
the expansion of the universe. This was the second stage in the development
of modern cosmology and was decisive in the establishment of the hot  big 
bang model as the standard cosmological model. 

\subsection{\label{Sect:Baryon-Asymmetry}Baryon Asymmetry in the Universe}

Although the standard hot big bang model is  successful in 
accounting for Hubble expansion, the residual microwave background radiation
and the abundances of light atomic nuclei there remain some serious difficulties
and one of these open questions is the observed baryon asymmetry 
in the universe today.
The essence of the baryon asymmetry problem is to understand why in the 
observable part of the universe the density of  baryons is many orders 
greater than the density of  antibaryons and why on the other hand the 
density of baryons is much less than the density of photons 
\cite{Kolb:1990}. The baryon asymmetry can be quantified by a 
quantity  known as the baryon
number $B$ of the universe and is defined  as the ratio of the baryon 
number density $n_{B}$ to the entropy density $s$. 
The baryon number density  $n_{B}$
is defined as the number density of baryons $n_{b}$ minus the number density
of antibaryons $n_{\overline b}$, so $n_{B}=n_{b}-n_{\overline b}$.  
The baryon number of the universe today, is estimated in the range
\cite{Kolb:1990,Kolb:ni} [See however in Section~\ref{Sect:Comments}]
\begin{equation}
B=\frac{n_{B}}{s} \sim(6-10)\times 10^{-11}~.
\end{equation}
There is a strong evidence of the asymmetry since it is well
known that the earth as well as all the objects on it consist of matter 
rather than antimatter. It is an experimental fact that there is no antimatter
on the earth and the outer space. Measurements of the cosmic rays 
emitted from the sun have proven the absence of antimatter in the solar system. 
Only an insignificant positron flux is contained in solar cosmic rays and it is 
mainly due to interactions of the primary solar cosmic rays with matter.

Astronomical observations show that at least our galaxy consists of usual 
matter, while an inconsiderable amount of antimatter is observed
and does not seem to indicate the presence of antimatter in the galaxy. For 
clusters 
of galaxies and their intergalactic gas one can derive limits for the amount of
antimatter present from the $\gamma$ flow expected from the $\pi{^0}$ decay,
since matter--antimatter annihilation would produce $\pi^{0}$ mesons and those 
decay into $\gamma$--rays. These hard photons have never been observed. It is not
definite if there are in our universe islands of antimatter separated by empty
space of regions containing normal matter. Electromagnetic radiation which is 
the main source of informations from the outer space cannot give a signal about 
its matter or antimatter origin.  

\subsection{\label{Sect:3-Conditions}Three Cornerstones of Baryogenesis}

In Friedman's  cosmology the excess of matter over antimatter is 
considered as one 
of the initial conditions and the baryonic asymmetry as one of the 
fundamental cosmological constants.
This philosophy changed in early seventies when Sakharov 
and others suggested that the observed asymmetry could be generated 
dynamically even though the universe had started in an initially symmetric 
state. Sakharov postulated three necessary conditions for this to happen:

\begin{verse}(a) There must be baryon number violating processes.\\
 (b)  Breaking of charge $(C)$ and combined $(CP)$ symmetries.\\
 (c)  These processes must go out of equilibrium sometime during the history
of the universe.
\end{verse}

The first condition is obvious since it is clear that the baryon number must 
be violated if the universe started its life in a baryon symmetric state and
then evolved into an asymmetric one. Many models predict baryonic asymmetry 
\cite{Dolgov:1991fr}, 
although baryon number violation has never been verified experimentally.
In Grand Unified Theories  (GUT)  baryon number violation occurs in the 
equations of motion of the theory. It proceeds with the exchange 
of very massive particles and takes place at an energy scale of the 
order of magnitude about ${10}^{16}\,\textrm{GeV}$. Supersymmetric  models also 
predict baryon non conserving processes in 
the energy range $10^{16}\,\textrm{GeV} $-- $10^{2}\,\textrm{GeV}$ depending upon the concrete 
model. 

In order to understand the necessity of the second condition we note that 
the baryon number of a state is odd under $C$ and combined $CP$
symmetries. That is, the baryon number of a state changes sign  under
charge conjugation $C$ and charge conjugation combined with parity $CP$. 
Thus if a state is either $C$ or $CP$ invariant, then the baryon number of
this state should be zero. If the universe begins its life with 
equal amounts of matter and antimatter, and without a 
preferred direction, then its initial state is both $C$ and
$CP$ invariant. Unless both $C$ and $CP$ are violated, the universe 
will remain $C$ and $CP$ invariant as it evolves to its final state, and
thus cannot develop a net baryon number  even if the baryon number is not
conserved. Therefore both $C$ and $CP$ violations are needed if a net 
excess of matter over antimatter is to be produced. In contrast to   
baryon number violation, which has only been predicted in various theoretical 
models but with no experimental evidence for it, $C$ and $CP$ violation have 
been verified experimentally in  neutral Kaon decays.

The necessity to satisfy the last Sakharov condition, that the universe 
must be out of thermal
equilibrium, is very important, since if an initially baryonic symmetric 
universe is in thermal equilibrium, then  
particle number densities are given by
\begin{equation}
n_{eq}=\frac{1}{\exp[(E-\mu)/T]\pm 1}~,
\end{equation}
where $E$ is the particle energy and  $\mu$ its chemical potential. If the 
charge is not conserved the corresponding chemical potential vanishes in 
equilibrium. $CPT$ invariance guarantees that particles and antiparticles have 
the same mass, therefore their number remains equal during the expansion and 
so no asymmetry occurs, regardless of $B$, $C$, or $CP$ violating processes.

\subsection{\label{Sect:Inflation}The Inflation Model}

The development of gauge theories and their influence to the cosmological 
questions led to the development of a new version of the
standard big bang model. The inflation model which was introduced by Guth
in 1981 \cite{Guth:1980zm} came to cure some of the pathological problems of
the hot universe theory. According to this model, soon after the 
beginning of the big bang the expansion rate of the universe increases 
exponentially, and due to this expansion  the universe supercools and 
all the physical processes are interrupted. Then the universe reheats 
again and continues its expansion with a slower rate. 

A detailed analysis of the inflation model
is given by Linde \cite{Linde:1990,Linde:ir}, while  important papers 
relevant to this model (and not only) are collected by Abbott and Pi 
\cite{Abbott:kb}. Although the inflation model is not the subject of this work, 
we should note here that it has already solved many of the 
problems that the hot big bang model possesses. On the 
other hand the inflation model imposes the constraint that the baryon generation
should have occurred after the exponential expansion, since the creation of a 
large amount of entropy during the inflation epoch would dilute any 
pre--existing baryon asymmetry. If one believes the inflation model then
baryogenesis should happen at rather lower scale energies  than those of
the GUT's scale. This is especially gratifying for present day 
investigations, since our technology makes only electroweak energy scales, not
GUT's scales, accessible to experimental tests.

%% file: chap2.tex
\section{\label{Sect:Gauge-Theories}Gauge Theories and Cosmology}

\subsection{Introduction}

The experimental success of the model of unified electroweak interactions and
the development of unified gauge theories of all the interactions at the 
beginning of the seventies opened a new era 
for the theories of the elementary particles. The creation 
and evolution of these theories have their roots on symmetry 
principles. Generalization of the ideas about symmetries generated the 
gauge principle
and the present belief that all particle interactions are dictated by the
so--called local gauge symmetries. This principle arises from the requirement
that all quantities which are conserved, are conserved locally and not merely
globally. 

Gauge invariant Lagrangians describing
particle interactions remain invariant under a set of local gauge
transformations. But invariance of the Lagrangian demands for the particles to 
be massless. The particles become massive due to the appearance of constant 
classical scalar fields $\phi$ over all space, through the so--called mechanism
of symmetry breaking.
 
The rapid development of particle physics not only led to a better 
understanding of particle interactions, but also to a significant progress in 
the theories of superdense matter, matter more than 80 orders of magnitude 
denser than the nuclear matter. It was Kirzhnits and Linde \cite{Kirzhnits:ut} and very soon  
others \cite{Dolan:qd,Weinberg:hy} who showed that 
the scalar fields 
which are responsible for the breaking of symmetry, could disappear at high 
enough temperatures. This means that at temperatures high enough phase 
transitions take place and after they have been completed, the symmetry 
is restored.

These theories are applicable in cosmology since superdense matter at very 
high 
energies and temperatures should appear at  very early stages of the evolution 
of the universe. According to the standard hot big bang model for an expanding 
universe, as the universe cools down these phase transitions could  
take place. Cosmological problems, as the baryon excess for example, could
find an answer in the framework of gauge theories.    
Recent development in this area suggest that the  baryogenesis problem could
find its answer in the standard model of electroweak interactions.  
This model possesses baryon number violation  as it was 
proved  by t'Hooft \cite{'tHooft:up,'tHooft:fv}, although the violation rate at 
zero temperatures seems 
to be too small. Violation of the $CP$ symmetry is provided by the 
Cabbibo--Kobayashi--Maskawa mixing between the quark families.
The condition  that the system must be out of equilibrium demands the 
electroweak phase transition to be of the first order. 

The remainder of this section is devoted on the formulation of 
the electroweak theory and an introduction to the theory of phase transitions
in gauge theories. In the next section we discuss the formulation of
the standard electroweak model as a gauge theory. The mechanism of symmetry 
breaking and the generation
of particle masses are presented in Sections~\ref{Sect:SSB} and \ref{Sect:Masses} respectively. The 
idea of symmetry restoration 
at high temperatures and a brief discussion of the cosmological phase 
transitions are contained in the last two Sections~\ref{Sect:Phase-Transitions} 
and \ref{Sect:Cosmological-Phase-Transitions}~.

\subsection{Electroweak Gauge Theory}

The successful formulation of Quantum Electrodynamics (QED) as a gauge theory
led to the idea of extending the gauge principle to the description of other 
interactions. Constructing a gauge theory we demand the Lagrangian to be 
locally invariant under a group of internal symmetries. Thus we introduce into 
the Lagrangian a number of vector fields equal to the number of the symmetry 
group generators. 
The gauge bosons self--interactions as well as the way they couple to  matter 
fields is completely determined by the gauge couplings. The QED Lagrangian for 
example is invariant under a set of local $U(1)$ transformations  
with generator $Q$, the electric charge operator. 
Invariant interactions are mediated by one vector field, the photon field. In 
Quantum Chromodynamics (QCD), the gauge theory of strong interactions, we 
introduce eight vector fields (the
gluon fields) into the Lagrangian and the theory is based on the 
group $SU(3)$, which has eight generators the Gell--Mann matrices. 

The Glashow--Salam--Weinberg  model which unifies the weak and electromagnetic 
interactions, known as the standard model of electroweak interactions, is 
based on the non--Abelian group $SU(2)_{L}\times U(1)_{Y}$ and has four 
generators. The weak interactions are mediated 
by three  vector bosons $A^{a}_{\mu},\; a=1, 2, 3$  which are accommodated in 
the adjoint representation of an $SU(2)$ group denoted by $SU(2)_{L}$ with 
generators $T_{a}=\tau_{a}/{2}$, where the $\tau_{a}$, $a=1, 2, 3$  are the 
Pauli matrices and the $T_{a}$ are referred as the weak isospin generators.
The subscript $L$ signifies that the $A_{\mu}^{a}$ couple in a parity 
violating 
way to the left  handed parts of the lepton matter fields, which transform as 
a doublet again  under the same $SU(2)_{L}$. 

To accomodate the electromagnetic interactions
we need another gauge boson $B_{\mu}$ and another group denoted 
by $U(1)_{Y}$. The generator of the $U(1)_{Y}$ group
is $Y/2$, where $Y$ is the weak  hypercharge defined through the so--called
Gell--Mann--Nishijima relation as
\begin{equation}
Q=T_{3}+Y/2 
\end{equation}
and $T_{3}$ is the third component of the  weak  isospin. The gauge 
coupling for the weak isospin group $SU(2)_{L}$ is  denoted by $g$ and 
that for weak hypercharge group $U(1)_{Y}$ as $g^{\prime}/2$. The kinetic 
terms into the Lagrangian for the two gauge fields are
\begin{equation}
\mathcal{L}_{gauge}=-\frac{1}{4}F^{a}_{\mu\nu}F^{a\mu\nu}-\frac{1}{4}B_{\mu\nu}
B^{\mu\nu}~,
\label{Eq:Gauge-Bosons}
\end{equation}
where the field strength tensor for the $SU(2)_{L}$ gauge field is given by
\begin{equation}
F^{a}_{\mu\nu}=\partial_{\mu}A^{a}_{\nu}-\partial_{\nu}A^{a}_{\mu}+
g\epsilon^{abc}A^{b}_{\mu}A_{\nu}^{c}~,
\end{equation}
while the field strength tensor of the $U(1)_{Y}$ gauge field has the form
\begin{equation}
B_{\mu\nu}=\partial_{\mu}B_{\nu}-\partial_{\nu}B_{\mu}~.
\end{equation}
The physical particles which mediate the weak and electromagnetic 
interactions,  the two charged $W^{+}$ and $W^{-}$, the neutral 
$Z^{0}$ and the photon $\gamma$ result as linear combinations of the 
fields $A^{a}_{\mu}$ and $B_{\mu}$.

The fermionic matter fields are present in the model grouped in three
families or generations which include  the $ e,\;\nu_{e},\; u,\; d$ known as 
first  
generation and other  two replications the $\mu,\;\nu_{\mu},\; c,\; s$ family
and the $\tau,\; \nu_{\tau},\; t,\; b$ family.
Violation of the parity is introduced into the model by putting the left--handed 
and right--handed fermions into different group representations. All
the left--handed fields are taken to transform as doublets under 
$SU(2)_{L}$, while the right--handed fermions transform as singlets. 

The left--handed quarks 
are introduced into the model grouped into $SU(2)$ doublets of the form
\begin{equation}
Q_{u}=\left(\begin{array}{l}
       u\\d^{\prime}
      \end{array}\right)
\qquad
%\hspace{1cm}
Q_{c}=\left(\begin{array}{c}
       c\\s^{\prime}
      \end{array}\right)
\qquad
%\hspace{1cm}
Q_{t}=\left(\begin{array}{r}
       t\\b^{\prime}
      \end{array}\right)~.
\end{equation}
The prime means that these are weak eigenstates, which are constructed
as linear combinations of the mass eigenstates. The matrix which connects
the two sets of eigenstates is referred as the 
Cabbibo--Kobayashi--Maskawa (CKM)
matrix. By convention the charge 2/3 quarks are unmixed, but the 
eigenstates
of the charge $-1/3$ quarks are related through the CKM matrix as
\begin{equation}
\left(\begin{array}{c} d^{'} \\ s^{'} \\ b^{'}
\end{array}\right)=K
\left(\begin{array}{c} d \\ s \\ b       
\end{array}\right)~.
\end{equation}
This mixing is very significant for violation of the $CP$ symmetry in the 
the standard electroweak model which we discuss in the next chapter. 
The right--handed quarks are $SU(2)$ singlets with hypercharge $Y=4/3$ for the
charge $Q=2/3$ quarks
$u_{R}, c_{R}$ and $t_{R}$ since the charge $Q=-1/3$ quarks $d^{\prime}_{R}, 
s^{\prime}_{R}$ and $b^{\prime}_{R}$ have hypercharge $Y=-2/3$.

In a similar way the  left-handed leptons are present in the model  grouped 
in $SU(2)$ doublets of the form
\begin{equation}
L_{e}=\left(\begin{array}{l}
       \nu_{e}\\e
      \end{array}\right)
\qquad
%\hspace{1cm}
L_{\mu}=\left(\begin{array}{c}
       \nu_{\mu}\\ \mu
      \end{array}\right)
\qquad
%\hspace{1cm}
L_{\tau}=\left(\begin{array}{r}
       \nu_{\tau}\\ \tau
      \end{array}\right)
\end{equation}
where the left--handed lepton states are $l_{L}=\frac{1}{2}(1-\gamma^{5})l$
and $l$ stands for $e,\; \mu,\; \tau$ and the corresponding neutrinos.   
The right--handed leptons are weak isospin singlets  of 
the form $l_{R}=\frac{1}{2}(1+\gamma^{5})l$. The $l$ now stands for the $e
,\; \mu,\; \tau$, since it is convenient to idealize the neutrinos as massless
and in this case right--handed neutrinos do not exist.  
In order to satisfy the Gell--Mann--Nishijima relation for the lepton charge
the left--handed leptons have hypercharge $Y_{L}=-1$, whilst
to the right--handed we assign hypercharge $Y_{R}=-2$~.

As a conclusion to the above discussion, the fermion part of the Lagrangian 
can be written as 
\begin{eqnarray}
\label{Eq:Fermions}
\mathcal{L}_f=&&\sum_{f_L} \overline f_{L}\gamma^{\mu}\left ( i\partial_
{\mu}-g\frac{\tau_{a}}{2}A^{a}_{\mu}-g^{\prime}\frac{Y}{2}B_{\mu}\right )f_{L} \nonumber \\
&&+\sum_{f_R} \overline f_{R}\gamma^{\mu}\left (i\partial_{\mu}-g^{\prime}\frac{Y}{2}
B_{\mu}\right )f_{R}~,
\end{eqnarray}
where the sum runs over all left-- and right--handed fermion fields. Addition 
of the gauge and fermion parts results in a Lagrangian density which
describes massless gauge boson fields interacting with massless fermionic
matter fields. 

Gauge boson mass terms which should be of the form
$\frac{1}{2}M_{B}^{2}B_{\mu}B^{\mu}$ plus similar terms for the $A^{i}_{\mu}$
are clearly not gauge invariant. In addition a fermion mass term should be 
of the form
\begin{equation}
-m\overline f f = -m(\overline f_{R}f_{L}+\overline f_{L}f_{R})~.
\end{equation}
Such a term manifestly breaks the gauge chiral invariance, since $f_{L}$
is a member of a $SU(2)_{L}$ doublet whilst $f_{R}$ is a singlet.
Thus addition of mass terms by hand into the Lagrangian destroys the
gauge invariance and the resulting theory is not renormalisable.
However, it is an experimental fact that the vector bosons which mediate 
the weak interactions and the fermions are massive particles. The answer to the
crucial problem of the mass generation without loosing the renormalisability
of the theory is the Higgs mechanism, which is discussed in the next section.

\subsection{\label{Sect:SSB}Spontaneous Symmetry Breaking}

The only known renormalisable theories involving vector bosons are gauge 
theories, but since mass terms in the Lagrangian spoil the gauge invariance    
it is impossible to find a renormalisable theory of massive vector bosons. 
However, it is possible for the Lagrangian and the equations of motion of a 
system to have a symmetry but the solutions of the equations to 
violate this symmetry. This phenomenon, where the solutions violate the 
symmetry of the equations, is known as spontaneous symmetry breaking (SSB). A 
well known example is the case of a ferromagnet, where the Hamiltonian of
the system is rotationally invariant, but the ground state of the system
has a non zero magnetization and it is clearly not invariant under rotations.
The analogous of the ground state of the ferromagnet in particles physics
is of course the vacuum. The Lagrangian of the theory is taken to be invariant
under an internal symmetry, but the vacuum state is not. The ground state  is
characterised by some scalar fields which are not invariant under the
symmetry transformations and develop a non zero vacuum expectation value. 

Consider the case of one scalar field with
quartic self interaction. The Lagrangian of the theory is given by
\begin {equation}
\mathcal{L}=\frac{1}{2}(\partial_{\mu}\phi)^{2}-\frac{1}{2}m^{2}\phi^{2}-
      \frac{1}{4}\lambda\phi^{4}~,
\end{equation}
where $m$ is the field mass and $\lambda$ its coupling constant. The 
potential energy density of this field at the classical level, which is
called the classical potential is given by 
\begin{equation}
V_{0}=\frac{1}{2}m^{2}\phi^{2}+\frac{1}{4}\lambda\phi^{4}~.
\label{Eq:V0}
\end{equation}
The Lagrangian remains invariant  under the 
the symmetry operation which replaces  $\phi$ by $-\phi$. 

\begin{figure}
\includegraphics[scale=0.47]{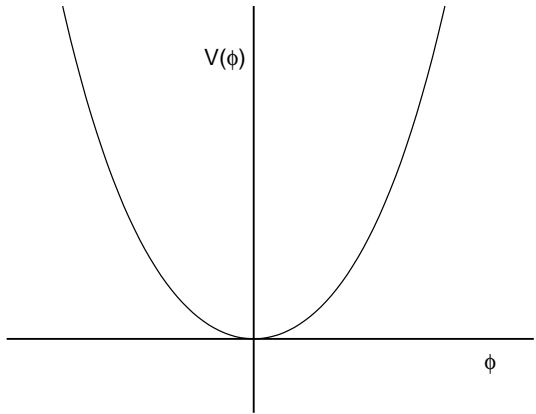} 
\caption{\label{Fig:21} The classical potential $V_{0}(\phi)$ before symmetry breaking.}
\end{figure}

Let us consider now two cases and first suppose that the mass squared of the 
field is positive, $m^{2}=\mu^{2}>0$. 
From  Eq.~(\ref{Eq:V0}) it is clear that the potential has a minimum, which
is the most favourable state of the system, for $\phi=0$. The minimum of 
the potential is shown in Fig.~\ref{Fig:21}~.

\begin{figure}
\includegraphics[scale=0.47]{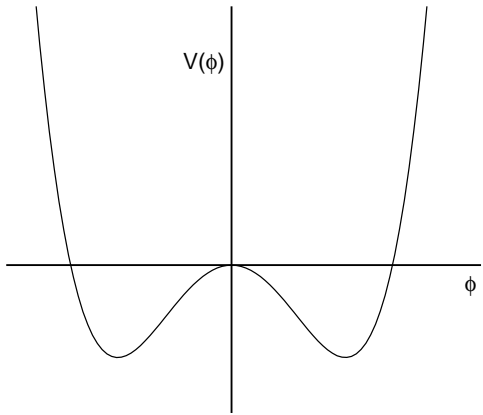} 
\caption{\label{Fig:22} The classical potential $V_{0}(\phi)$ after symmetry breaking.}
\end{figure}

But in the case when the mass squared is negative, that is 
$m^ {2}=-\mu^{2}<0$, we observe a very different situation . The potential 
energy has two minima which occur at 
$\phi=\sigma=\pm\mu/\sqrt{\lambda}$ 
and  now there are  two degenerate 
ground states. Once we choose one of them, the ground state does not share
the symmetry with the Lagrangian.This situation is illustrated in Fig.~\ref{Fig:22}.
The value of the curvature of the potential near the extrema describes the 
effective mass squared of the scalar fluctuations. Prior the symmetry 
breaking  the mass term of the field is negative. 
After breaking of symmetry
the mass term becomes positive
\begin{equation}
m^{2}(\phi=\sigma)=\frac{d^{2}V}{d\phi^{2}}\bigg |_{\phi=\sigma}
=3\lambda \sigma^{2}-\mu^{2}=2\mu^{2}>0~.
\end{equation}    

In determining masses and interactions one must perturb about the stable 
vacuum by shifting the field $\phi$ as $\phi(x)=\sigma+\eta(x)$~. Inserting 
that into the Lagrangian, the mass term appears with the correct sign.
The above example serves as a useful 
introduction to symmetry 
restoration at high temperatures, which we discuss in Section~\ref{Sect:Phase-Transitions}. 

\subsection{\label{Sect:Masses}Boson and Fermion Masses}

In order to understand the problem of mass generation for  fermions
and vector bosons we review  here the
well known Higgs model. In the simple scalar model, which we examined earlier,
the mass term after symmetry breaking appeared with the correct sign. In the 
Higgs model insertion of the scalar field into the Lagrangian results
in the generation of masses for  the particles which interact with. The 
Lagrangian of this model exhibits a $U(1)$ gauge symmetry and is given by
\begin{eqnarray}
\mathcal{L}&=&(\partial^{\mu}+igA^{\mu})\phi^{*}(\partial_{\mu}-igA_{\mu})\phi
-\frac{1}{4}F_{\mu\nu} F^{\mu\nu} \nonumber \\
& &-\mu^{2}\phi^{*}\phi-\lambda(\phi^{*}\phi)^{2}~.
\end{eqnarray}
For $\mu^{2}>0$ the Lagrangian, apart from the scalar self interaction term, 
is that of scalar QED. However, in the case where  $\mu^{2}<0$, shifting the 
field to its true ground state as 
\begin{equation}
\phi(x)=\frac{1}{\sqrt{2}}[\sigma+\eta(x)+i\xi(x)]
\end{equation}
and inserting that into the Lagrangian, the latter becomes 
\begin{eqnarray}  
\mathcal{L}^{\prime}&=&\frac{1}{2}(\partial_{\mu}\xi)^{2}+\frac{1}{2}(\partial_{\mu}
\eta)^{2}- \frac{1}{4}F_{\mu\nu}F^{\mu\nu} \nonumber \\
&&-\lambda \sigma^{2}\eta^{2}+\frac{1}{2}g^{2}\sigma^{2}(A_{\mu})^{2}
-g\sigma A_{\mu}\partial_{\mu}\xi
\end{eqnarray}    
where for simplicity we have omitted the interaction terms. 
The particle spectrum in the new Lagrangian appears to be composed by a 
massless scalar, the so-called Goldstone boson, a massive scalar 
particle  $\eta$ with mass $m_{\eta}=\sqrt{2\lambda \sigma^{2}}$ and 
a massive vector 
$A^{\mu}$ with mass $m_{A}=g\sigma$.

One can use the freedom of a gauge transformation to eliminate the 
unphysical Goldstone mode. Then the Lagrangian looks like
\begin{eqnarray}  
\mathcal{L}^{\prime\prime}&=&\frac{1}{2}(\partial_{\mu} h)^{2}
-\lambda \sigma^{2} h^{2}+\frac{1}{2}g^{2}\sigma^{2}A_{\mu}^{2}-\lambda \sigma
 h^{3}-\frac{1}{4}\lambda h^{4}\nonumber\\
                     &&+\mbox{}\frac{1}{2}g^{2}A_{\mu}^{2}h^{2}
+\sigma g^{2}A_{\mu}^{2}h-\frac{1}{4}F_{\mu\nu}F^{\mu\nu}.
\end{eqnarray}    
The Goldstone boson no longer appears in the theory and the Lagrangian 
describes 
just two interacting massive particles. The particle spectrum consists of
a massive vector boson $A_{\mu}$ and a massive scalar $h$, which is known as 
a Higgs particle. 

The mechanism we have presented above, is known as the Higgs mechanism. 
An overview of the model can be found for 
example in Halzen and Martin \cite{Halzen:mc} or Cheng 
and Li \cite{Cheng:bj}. The Higgs mechanism  can be generalized to non Abelian gauge 
theories and in 
particular to  the case of the standard model of electroweak interactions. 
We present how, due to spontaneous breaking of symmetry, the fermions and 
vector bosons acquire their masses. In the case of  the so--called 
minimal standard model, one complex $SU(2)$ doublet of scalar fields, with
hypercharge $Y=1$ is introduced,
\begin{equation}
\Phi=\left(\begin{array}{c}  
        \phi^{+}\\\phi^{0}
      \end{array}\right)=\frac{1}{\sqrt 2}\left(\begin{array}{c}  
        \phi_{1}+i\phi_{2}\\\phi_{3}+i\phi_{4}
      \end{array}\right)~,            
\end{equation}      
where $\phi,\;\;i=1,2,3,4$ are real fields.
The  building block of the Lagrangian containing the scalar doublet looks like 
\begin{equation}
\mathcal{L}_{\phi}=(D^{\mu}\Phi)^{\dag}(D_{\mu}\Phi)-V(\Phi)~,
\label{Eq:Scalar}
\end{equation}
where the covariant derivative is given by
\begin{equation}
D_{\mu}=\partial_{\mu}-i\frac{g}{2}\tau^{a}A^{a}_{\mu}-i\frac{g^{\prime}}{2}
B_{\mu}
\end{equation}
and the most general gauge--invariant renormalisable potential is of the form
\begin{equation}
V(\Phi)=-\mu^{2}\Phi^{\dag}\Phi+\lambda(\Phi^{\dag}\Phi)^{2}~,
\end{equation}
with the mass $\mu^{2}$ being positive. Without loss of generality the minimum
of the field $\Phi$ can be chosen at
\begin{equation}
\Phi=\frac{1}{\sqrt 2}\left(\begin{array}{c}  
       0\\\sigma
      \end{array}\right)~,
\end{equation}
where $\sigma^{2}=\mu^{2}/\lambda$ and since we are interested in the physical
particle spectrum the unitary gauge is an appropriate choice. Just as in the 
case of the Higgs
model, after spontaneous breaking of symmetry 
the kinetic term of the scalar part of the Lagrangian $(D^{\mu}\Phi)^{\dag}
(D_{\mu}\Phi)$  gives rise to the vector boson masses. 
The two charged vector bosons are defined by the combination
\begin{equation}
W^{\pm}_{\mu}=\frac{1}{\sqrt {2}}(A^{1}_{\mu}\pm iA^{2}_{\mu})
\end{equation}
with a classical mass
\begin{equation}
m_{W}=\frac{1}{2}g\sigma~,
\end{equation}
where $\sigma$ is the vacuum expectation value of the Higgs field. The neutral vector 
boson field appears as  the combination
\begin{equation}
Z_{\mu}=\frac{g^{\prime}B_{\mu}-g A^{3}_{\mu}}{\sqrt{g^{2}+g^{\prime^{2}}}}
\end{equation}
and its  classical mass is given by
\begin{equation}
m_{Z}=\frac{1}{2}(g^{2}+g^{\prime^{2}})^{1/2}\sigma.
\end{equation}
The photon field remain massless $m_{A}=0$ and is given by the 
combination  
\begin{equation}
A_{\mu}=\frac{g^{\prime}B_{\mu}+g A^{3}_{\mu}}{\sqrt{g^{2}+g^{\prime^{2}}}}~.
\end{equation}

The generation of the fermion masses can be done in a similar way. For this
we introduce  into the Lagrangian Yukawa coupling terms between scalars and 
fermions of the general form
\begin{equation}
\mathcal {L}_{Y}=g_{e}\overline L_{e}\Phi e_{R} + g_{d}\overline Q_{u}\Phi d_{R}
 + g_{u}\overline Q_{u}\tilde\Phi u_{R}+h.c~.
\label{Eq:Yukawa}         
\end{equation}
for all the fermion doublets. In order to generate the quark masses for  the upper components   
of a quark doublet  we have constructed a new Higgs doublet
from  $\Phi$
\begin{equation}
\tilde\Phi=-i\tau_{2}\Phi^{\dagger}=\left(\begin{array}{c}  
        -\overline\phi^{0}\\\phi^{-}
      \end{array}\right)~,
\end{equation}      
which transforms identically to $\Phi$, but has opposite weak 
hypercharge $Y=-1$. After symmetry breaking the neutrinos remain massless, while the quarks and 
the lower components of the lepton 
families acquire masses of the form
\begin{equation}
m_{f}=\frac{g_{f}\sigma}{\sqrt{2}}~,
\end{equation}
where $g_{f}$ is the Yukawa coupling of fermion $f$ to the scalar doublet
$\Phi$.

The final Lagrangian of the standard electroweak model appears as  a sum of 
the gauge boson term  Eq.~(\ref{Eq:Gauge-Bosons}), the fermion term Eq.~(\ref{Eq:Fermions}), the above 
given scalar part Eq.~(\ref{Eq:Scalar}) and the Yukawa interaction terms Eq.~(\ref{Eq:Yukawa}). In 
addition to those terms a gauge fixing term must
be included into the Lagrangian. The unitary gauge which we had chosen
earlier is convenient and displays clearly the particle spectrum, but it is not
the most appropriate when one calculates quantities involving the propagators 
of the vector bosons. Instead of the unitary gauge, a class of gauges like the
so--called $R_{\xi}$ gauges can be chosen. In this case the gauge
fixing term has the form
\begin{equation}
\mathcal{L}_{GF}=\frac{1}{2\xi}(\partial^{\mu}A^{a}_{\mu}-\frac{1}{2}\xi
gv\chi^{a})^{2}-\frac{1}{2\xi}(\partial^{\mu}B_{\mu}-\frac{1}{2}\xi 
g^{\prime^{2}}v\chi^{a})^{2}~.
\end{equation}   
Different values of $\xi$ correspond to different gauges. One 
choice is
the Landau gauge ($\xi=0$), which although has a bit complicated 
propagators, is very convenient in calculations involving scalar potentials  
because 
the coupling of unphysical scalars (Goldstone bosons) to physical scalars is 
zero. On the other hand in calculations involving the effective potential
(which we discuss in Section~\ref{Sect:Veff}) the Landau gauge is used 
\cite{Dolan:qd,Sher:1988mj}, mainly because
of the vanishing interaction between physical scalars and ghosts (necessary
ingredients of non abelian gauge theories).

\subsection{\label{Sect:Phase-Transitions}Phase Transitions in Gauge Theories}   

As it was referred in the introduction, a raising of the temperature 
results in 
the disappearance of the classical field which is responsible for the breaking
of symmetry and this corresponds to a phase transition 
\cite{Linde:1990,Linde:px}. Suppose we deal with one scalar 
field $\phi$. At zero temperature, the 
location of the minimum of the potential $V_{0}(\phi)$
describes the true ground state, the vacuum of the theory. At a finite 
temperature $T$ the equilibrium state of this field   is governed
by the location of the minimum of the free energy density $F(\phi,T)\equiv
V(\phi,T)$,  which is equal to $V_{0}(\phi)$ at zero temperature. 
The  contribution to the free energy  density from ultrarelativistic scalar 
particles  of mass $m$ at a temperature $T$ which is much larger than 
the particle mass has the form
\begin{equation}
\Delta F=\Delta V(\phi, T)=-\frac{\pi^{2}}{90}T^{4}+\frac{m^{2}T^{2}}{24}
         +O(\frac{m}{T})~.
\end{equation}
The resulting potential energy density at finite temperature, which 
is known as the effective potential $V(\phi, T)$, is the sum of 
the classical potential plus the above contribution.
The appearance of these contributions to the classical potential is discussed
in some detail in Section~\ref{Sect:Veff}, where the theory of the effective
potential is explored.                    
As we have shown earlier the field dependent particle mass is given by
\begin{equation}
m^{2}(\phi)=\frac{d^{2}V}{d\phi^{2}}=3\lambda\phi^{2}-\mu^{2}~,
\end{equation}    
so the full expression of the temperature dependent potential, at the limit 
of temperatures large enough compared to the particle masses looks like
\begin {equation}
V(\phi, T)=-\frac{\mu^{2}\phi^{2}}{2}+\frac{\lambda\phi^{4}}{4}  
           +\frac{\lambda T^{2}\phi^{2}}{8}-\frac{\pi^{2}T^{4}}{90}
           -\frac{\mu^{2}T^{2}}{24}~.
\end{equation}
At zero temperature the effective potential has a local maximum at $\phi=0$
and an absolute minimum at $\phi=\sigma\neq 0$, but as the 
the temperature increases, the energy difference between the minimum of 
the potential at $\phi\neq 0$ and the local maximum at $\phi= 0$ decreases.
At a critical temperature, which is given by 
\begin{equation}
T_{c}=2\frac{\mu}{\sqrt{\lambda}}~,
\end{equation}
this energy difference disappears and the only minimum of $V(\phi,T)$ is 
the one at $\phi=0$.

\begin{figure}
\includegraphics[scale=0.47]{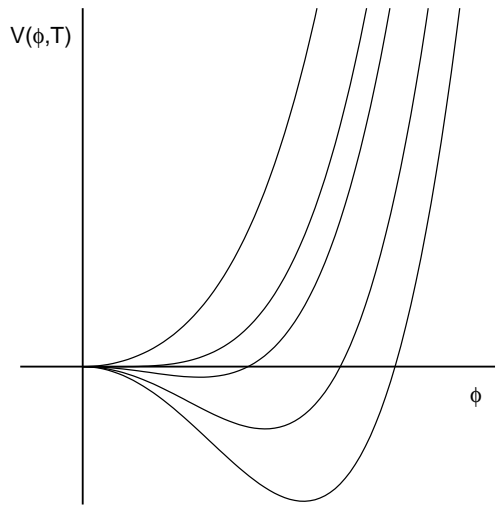} 
\caption{\label{Fig:23}Evolution of the effective potential $V(\phi, T)$ in the case 
of a second order phase transition.}  
\end{figure}

This corresponds to a  phase transition which takes 
place from a state with broken symmetry to a symmetric state and 
the particles appear to be  massless again. The behaviour of 
the effective potential for several temperatures 
is given in the next qualitative picture in Fig.~\ref{Fig:23}~. 

Any phase transition is described in terms of an order parameter which
distinguishes the various phases. There are two types of phase transitions.  
In phase transitions of the first type the order parameter jumps
discontinuously from its value in the first phase to that in the second. A
phase transition of this type is called a first order phase transition. 
Phase transitions  where the order parameter changes continuously are 
known as second order phase transitions. In our previous example the order 
parameter of the transition is the classical field. During the transition
the value of the classical field at the minimum decreases 
continuously to zero and the transition is of the second order. 

A more complicated case appears in theories with
more than one coupling constants, as for example the Higgs model, which we have
discussed in the previous section. 
In the Higgs model when the temperature is high enough as  compared to the 
particle masses and if one assumes that $\lambda\sim g^{2}$, temperature 
induced effects add a term to the classical potential, so it looks like
\cite{Kirzhnits:ut,Linde:1990,Linde:px}
\begin{equation}
V(\phi, T)=-\frac{1}{2}\mu^{2}\phi^{2}+\frac{1}{4}\lambda\phi^{4}
+\frac{4\lambda+3g^{2}}{24}T{^2}\phi^{2}.
\end{equation}
The phase transition takes place at a  critical temperature which is given by
\begin{equation}
T^{2}_{c}=\frac{12\mu^{2}}{4\lambda+3g^{2}}
\end{equation}
and we deal again with a second order phase transition, since the field 
depends on the temperature continuously. 

\begin{figure}
\includegraphics[scale=0.47]{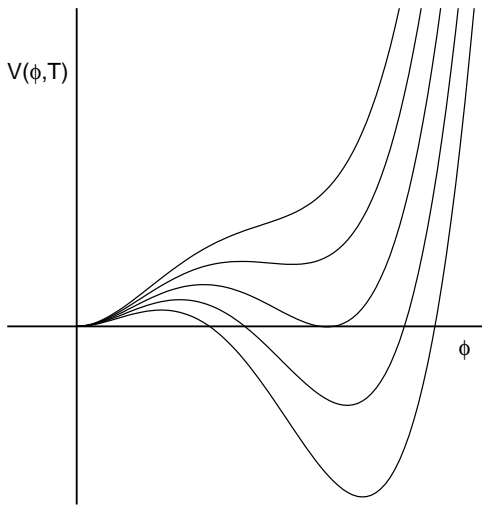} 
\caption{\label{Fig:24}Evolution of the effective potential $V(\phi, T)$ in the case 
of a first order phase transition.}
\end{figure}

However, a different situation appears  if we  consider the case
$\lambda\leq g^{4}$ since  the  mass  of the vector particles, which
is given by $m(\phi)=g\phi$, can no
longer be neglected compared to temperature and
their contribution to the effective potential  result in the form 
\begin {equation}
\Delta V(\phi, T)=-\frac {3m^{3}(\phi)T}{12\pi}
\end{equation}
plus field independent terms which we omit. 
The extrema of the effective 
potential correspond to the solutions of the equation
\begin{equation}
\frac{dV(\phi, T)}{d\phi}=\phi(\lambda\phi^{2}-\mu^{2}+\frac{4\lambda+3g^{2}}
{12}T^{2}-\frac{3g^{3}}{4\pi}T\phi)=0.
\end{equation}
The behaviour of the effective potential as a function of the field 
for several temperatures is given in Fig.~\ref{Fig:24}.

As we can observe due to the appearance of the cubic term in the field the 
effective potential has now three extrema. There are two local minima at 
$\phi=v_{1}$ and at $\phi =0$ and a  local maximum           
at $\phi=v_{2}$. The state which corresponds to $\phi=v_{1}$ is stable at 
low temperatures but it becomes metastable above the critical temperature. 
When the system reaches the critical temperature, the potential has two 
degenerate local minima, one giving the broken symmetry phase $\phi\not=0$ 
and the other the symmetric phase $\phi=0$. At the critical temperature as
the system cools down a phase transition takes place from the symmetric 
phase to the state with broken symmetry. This phase transition is of the 
first order, since the field depends on the temperature  discontinuously.
In  particular when $3g^{2}/16\pi^{2}\leq\lambda\leq g^{4}$, the critical 
temperature in the Higgs model is given by
\begin{equation}
T_{c}=\left(\frac{15\lambda}{2\pi^{2}}\right)^{1/4}\mu.
\end{equation}

\subsection{\label{Sect:Cosmological-Phase-Transitions}Cosmological Phase Transitions}

Phase transitions like the ones we presented above could have occurred during 
the expansion of the universe.  
According to the standard hot universe theory, the universe started its life
from a state with enormous density and temperature and with symmetry 
restoration between all the interactions. At the beginning of the expansion 
all the particles appear to be massless. As the universe cools down
phase transitions take place breaking the symmetry between the interactions 
and result in the generation of the massive particles we observe today.
It is believed  that the universe started from an initial state at 
a temperature which was at least of the order of the Planck scale about
$T\sim 10^{19}\,\textrm{GeV}$. During the expansion three such phase transitions
could have occurred. The first phase transition, known as the GUT's phase 
transition, could have
happened at a temperature of the order 
$T_{c_{1}} \sim 10^{14}-10^{15}\,\textrm{GeV}$. Through this phase
transition a scalar field $\phi \sim 10^{15}\,\textrm{GeV}$ is generated, which
breaks the symmetry between the strong and electroweak interactions.
As the temperature decreases further, at a temperature of the order of
magnitude $T_{c_{2}} \sim 10^{2}\,\textrm{GeV}$,  
a second phase transition can  take place and the symmetry between the weak 
and electromagnetic interactions breaks. This is called the electroweak
phase transition and is the main subject
of this work. The nature of the electroweak phase transition will be
investigated in detail in Section~\ref{Sect:EWPT}.  Finally, at a temperature
$T \sim 10^{2}\,\textrm{MeV}$, the QCD phase transition takes place
which breaks the chiral invariance of the strong interactions.

%% file: chap3.tex
\section{\label{Sect:BCP-Violation}Baryon Number and $CP$ Violation}

\subsection{Introduction}

As it was stated in Section~\ref{Sect:Baryogenesis-Problem} the Sakharov conditions demand that
baryon number and $CP$  must be violated, in order to produce a net 
baryon number. Since the work of t'Hooft \cite{'tHooft:fv},
baryon number violation
has been shown to occur in the standard electroweak model through the axial 
anomaly, although the rate of 
anomalous baryon number non conserving processes are exponentially suppressed
at zero temperatures. As the temperature increases the rate of baryon number
violation  increases and is no more negligible at temperatures close to the 
electroweak scale.
The electroweak theory violates charge conjugation $C$, while 
charge conjugation combined with parity ($CP$) 
is violated through
the quark Yukawa couplings, although this $CP$ violation appears to be too 
small to generate the required asymmetry.

Our aim is to sketch how baryon number violation and $CP$ 
violating effects take place within the physics of the standard electroweak
model. Baryon number violation at zero and finite temperature is discussed 
in Section~\ref{Sect:B-Violation}, while the final 
Section~\ref{Sect:CP-Violation} is devoted to $CP$ violation as it appears in the standard 
electroweak model. 

\subsection{\label{Sect:B-Violation}Baryon Number Violation}

\subsubsection{Zero Temperature}

In the standard electroweak model baryon number $J^{\mu}_{B}$ and
lepton number $J_{L}^{\mu}$ currents are exactly conserved at classical
level. However quantization of the theory leads to the appearance of anomalous
axial currents \cite{'tHooft:fv}. In what follows 
we ignore electromagnetic effects, which
mathematically correspond  to setting the Weinberg angle to zero
\cite{McLerran:1992sg,Turok:1992ar,Turok:1990in,Shaposhnikov:1991pd}. It is 
as if by quantum corrections the $W$ and $Z$ boson fields have acquired 
an indeterminate baryon plus lepton number.
 
Violation of the baryon number and lepton number currents  
in the standard electroweak model comes from two
crucial facts. The first is that  due to the nature of the
electroweak interactions the gauge boson fields couple differently to the 
left--handed and right--handed fermion fields. When a fermion couples to a 
gauge boson the corresponding axial current is anomalous
\begin{equation}
\label{Eq:Axial-Current}
\partial_{\mu}J^{\mu}_{B}=N_{F}\frac{g^{2}}{32\pi^{2}}F^{a}_{\mu\nu}
\tilde F^{a\mu\nu}~,                                    
\end{equation} 
where $N_{F}$ is the number of fermion generations and $g$ is the
$SU(2)_{L}$ coupling constant. The field 
strength tensor of the $SU(2)_{L}$ group is given by
\begin{equation}
F_{\mu\nu}^{a}=\partial_{\mu}A_{\nu}^{a}-\partial_{\nu}A_{\mu}^{a}    
-g\epsilon^{abc}A^{a}_{\mu}a^{c}_{\nu}~,
\end{equation}
and $\tilde F^{\mu\nu}$ is related to $F^{\mu\nu}$ through
\begin{equation}
\tilde F_{\mu\nu}\equiv\frac{1}{2}\epsilon^{\mu\nu\rho\sigma}
F_{\rho\sigma}~.
\end{equation}
In the above equation  $F_{\rho\sigma}=F^{a}_{\rho\sigma}T^{a}$ where 
$T^{a}$ are the generators of the gauge group algebra.
Calculation of the anomaly for the lepton number  current gives an identical
result, so $B-L$ is conserved but $B+L$ is not.

It is very significant that the term on the right hand side of the Eq.~(\ref{Eq:Axial-Current})
may be written as four--divergence  of  a new current, the so--called
Chern--Simons current 
\begin{equation}
\label{Eq:k-mu}
K^{\mu}=\epsilon^{\mu\nu\rho\sigma}(F^{a}_{\nu\rho}A^{a}_{\sigma}
-\frac{2}{3}\epsilon^{abc}A^{a}_{\nu}A^{b}_{\rho}A^{c}_{\sigma})
\end{equation}        
since one can find that
\begin{equation}
\partial_{\mu}K^{\mu}=F^{a}_{\mu\nu}\tilde F^{a\mu\nu}~.
\end{equation}

The second crucial fact is that 
the standard electroweak model with an $SU(2)$ gauge group and a Higgs field 
possesses a non--trivial vacuum structure. In addition to the usual vacuum 
configuration $A_{i}=A^{a}_{\mu}\tau^{a}=0$ and $\phi=\phi_{0}$, there is 
an infinite number of pure gauge transformations of the form 
\begin{eqnarray}   
\phi^{\prime} &=&T(\Omega)\phi_{0}\nonumber\\
A^{\prime}_{i}&=&\frac{2i}{g}\partial_{i}\Omega\Omega^{-1}~,
\end{eqnarray}
where  $\Omega(x)$ is a $2\times 2$ unitary matrix  and $T(\Omega)$ an
arbitrary rotation matrix and under these transformations the Lagrangian 
of the model remains invariant.

Physics imposes constraints, so one can find a non-local variable but 
locally gauge invariant quantity defined as
\begin{equation}
N_{CS}=\frac{g^{2}}{32\pi^{2}}\epsilon^{ijk}\int d^{3}x(F^{a}_{ij}
A^{a}_{k}
-\frac{2}{3}\epsilon^{abc}A^{a}_{i}A^{b}_{j}A^{c}_{k})~,
\end{equation}
which characterizes these degenerate vacua and it known  as 
the  Chern-Simons number. 
Under ``small'' gauge transformations 
the integrand of $N_{CS}$ changes by a quantity which is a total  divergence.
But under a ``large'' gauge transformation  $\Omega\rightarrow 1$, as the 
fields approach zero at the spatial infinity, the Chern--Simons 
number $N_{CS}$ is shifted by an integer number $n$ which is
called the winding number. The winding number is given by 
\begin{equation}
n[\Omega]=\frac{1}{24\pi^{2}}\int d^{3}x\epsilon^{abc}Tr(\Omega\partial_{a}
\Omega^{-1}\Omega\partial_{b}\Omega^{-1}\Omega\partial_{c}\Omega^{-1})~,
\end{equation}   
which is a non--trivial topological integral and it can be any real number but
for vacuum configurations of the field it is an integer. The winding number
characterizes the degree of mapping of the space compactified as a 3--sphere
onto the group $SU(2)$.

\begin{figure}
\includegraphics[scale=0.52]{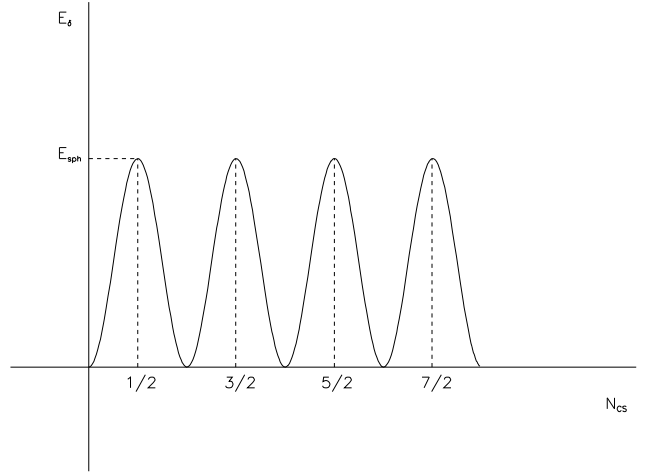} 
\caption{\label{Fig:31}The potential energy of the $SU(2)$ gauge field as a function of the 
Chern--Simons number. The minima correspond to configurations with zero gauge
field energy but different baryon number}
\end{figure}

The classical potential energy $E$ of the $SU(2)$ gauge field as a 
function of 
the Chern--Simons number $N_{CS}$ is multiply periodic and it can be 
schematically represented as in Fig.~\ref{Fig:31}. 

Between the degenerate vacua 
there exist unstable, time--independent solutions of the field equations called
sphalerons. The sphalerons were the first  introduced by  
Klinkhamer and Manton \cite{Klinkhamer:1984di} and they have Chern--Simons numbers 
$N_{CS}=1/2, 3/2....$ 
At these saddle points the height of the barrier separating topologically
distinct vacua has been estimated in the range
\begin{equation}
E_{sph}=\frac{2M_{W}}{a_{w}}B(\lambda/g^{2})=8-14\,\textrm{GeV}
\end{equation}
for $\lambda$ varying from zero  to infinity.

It is very significant that this non--trivial vacuum topology is related to the
anomaly equation. The total baryon number is given by
\begin{equation}
B=\int d^{3}xJ^{0}_{B}
\end{equation}  
and it can be shown that the baryon number change
during any time interval is related to the Chern--Simons number through
the equation
\begin{equation}
\Delta B=N_{F}\left[N_{CS}(t_{2})-N_{CS}(t_{1})\right]~,
\end{equation}
since the time component of $K^{\mu}$ in Eq.~(\ref{Eq:k-mu}) is simply the 
Chern--Simons number $N_{CS}$. Therefore if the baryon number is not to be
conserved it is related to a change in the Chern--Simons number of the $SU(2)$
gauge vacuum. 

At zero temperatures or low densities a transition  from 
one vacuum configuration to another can be achieved by quantum 
tunneling. In the computation of the transition rates, a classical solution
of the equations of motion can be used which is called an instanton. The 
instantons correspond to the classical solutions used in the WKB method of
quantum mechanics \cite{McLerran:1992sg,Coleman:1988,Weinberg:hc}. The transition 
rates have been computed by 
t'Hooft \cite{'tHooft:up,'tHooft:fv}, but as it was shown these are suppressed by a 
semiclassical factor 
$\exp(-4\pi/a_{W})\sim 10^{-150}$. At zero temperature therefore baryon number
non conservation is negligible. This of course is a consequence of the very
large energetic barrier $(\sim10\,\textrm{GeV})$ separating the vacua of different
baryon number due to the sphaleron but fortunately this picture is completely
different at high temperatures. 

\subsubsection{Nonzero Temperature}
  
At finite temperatures thermal fluctuations with an energy 
higher than the height of the barrier between the two degenerate minima 
can classically traverse the maximum and enter the new minimum, leading 
to unsuppressed baryon number violation. The rate of penetration of the 
sphaleron barrier is then given by the Boltzmann factor associated with
the formation of the sphaleron
\begin{equation}
\Gamma=(\alpha_{W}T)^{4}\exp[-E_{sph}(T)/T]~,
\end{equation}
where the sphaleron energy at temperature $T$ is given by
\begin{equation}
E_{sph}\sim B \frac {M_{W}(T)}{\alpha_{W}}~.
\end{equation}
The factor $B$ is  obtained numerically by solving the field equations. Such
a numerical solution is still lacking, however we can expect that for 
temperatures larger than the $W$ boson mass, the baryon  violation rate
should behave like
\begin{equation}
\Gamma\sim \alpha_{W}T^{4}~,
\end{equation}
due to the vanishing of masses at high temperatures. This of course leads
to a completely unsuppressed rate of baryon number violation. In particular 
this means that any pre--existing baryon asymmetry  generated during the
GUTs phase transition, would have disappeared until the time when the
EW phase transition takes place. However these processes only violate
$B+L$ since as we have seen $B-L$ does not have an anomaly and is exactly
conserved in the standard model. Therefore a $B-L$ excess produced in the 
early universe will not be washed  away. 

It is very significant the role played by the sphaleron after the completion 
of the electroweak phase transition \cite{Turok:1992ar,Turok:1990in}. The sphaleron mass, which is 
proportional to the Higgs vacuum expectation value $\sigma(T_{c})$, must be
large enough to strongly suppress baryon number violating 
processes. Otherwise, these processes may erase any asymmetry produced during
the phase transition. For baryons to survive we need the sphaleron rate to be
much less than the expansion rate of the universe.
These considerations impose the constraint 
\begin{equation}
\sigma(T_{c})/T_{c}\geq 1~,
\end{equation}
where $T_{c}$ is the temperature immediately  after the transition and can
also be used in order  one to obtain an upper bound for the Higgs mass, which
has found to be $m_{H}\leq 45\,\textrm{GeV}$ 
\cite{Shaposhnikov:tw,Shaposhnikov:1987pf,Bochkarev:1990gb}. 

\subsection{\label{Sect:CP-Violation}$CP$ Violation}

It has been well understood that $C$ and $CP$ violation is a crucial 
feature of any
theory that attempts to explain the observed asymmetry between matter and
antimatter in the universe starting from initially symmetric conditions. 
$CP$ symmetry requires that the partial rates of $CP$--conjugate processes
be equal, therefore for any process that violates the baryon number there 
would be a $CP$--conjugate process of equal rate and no asymmetry could be 
generated. So far the Kaon system has been the only laboratory for 
the observation of  $CP$ violation.  

Violation of $C$ symmetry is a generic feature in the standard model of
electroweak interactions, since 
due to the chiral nature of the $SU(2)$ gauge interaction, left and right
chiralities of quarks have different interactions, so the Lagrangian of 
standard model has both $C$--even and $C$--odd pieces. 

The standard model with two fermion families and a single Higgs doublet is 
a $CP$ invariant theory, but addition of the third generation violates the 
$CP$ invariance of the Lagrangian. As we referred in the second 
chapter on the formulation of the minimal standard model with one Higgs
doublet, $CP$ violation in the model originates from  the 
Cabbibo--Kobayashi--Maskawa (CKM) matrix which relates the quark mass
eigenstates to the weak eigenstates. 

The Yukawa term of the standard model Lagrangian involving the quark
interactions with the scalar doublet was given in Eq.~(\ref{Eq:Yukawa}), but in 
terms of fields which are eigenstates of the weak interactions, can be 
written in the form
\begin{equation}
{\cal L}_{Y}=\frac{g_{W}
}{\sqrt{2}M_{W}}=\left[\overline Q_{L}KM_{d}d_{R}\Phi   
\overline Q_{L}KM_{u}u_{R}\tilde\Phi+\;h.c.\right]~,
\end{equation}
where $K$ is the CKM matrix which relates the two set of eigenstates and 
$M_{u}$, $M_{d}$ are the diagonal mass matrices of the quarks with 
charges 2/3 and
$-1/3$ respectively. The specific form of the CKM matrix can be found  in 
\cite{Hikasa:je}.  

For three fermion generations the CKM matrix must be $3\times 3$ and 
unitary and this constraint provides relationships between its elements,
so finally its parameterisation can be done through three rotation angles and  
there can be only a physically 
meaningful phase which leads to observable  $CP$ violation effects
\cite{Farrar:hn,Shaposhnikov:1991cu,Jarlskog:jb}.  
This $CP$ violation can be quantified through the combination \cite{Jarlskog:1985ht}
\begin{eqnarray}
\label{Eq:Delta-CP}
\Delta_{CP} & = & \sin\,\theta_{12}\;\sin\,\theta_{23}\;\sin\,\theta_{13}\;\sin\,\delta_{CP}\nonumber \\   
&& \times(m_{t}^{2}-m_{c}^{2})(m_{t}^{2}-m_{u}^{2})(m_{c}^{2}-m_{u}^{2})\nonumber \\ 
&& \times(m_{d}^{2}-m_{s}^{2})(m_{b}^{2}-m_{d}^{2})(m_{s}^{2}-m_{d}^{2})~,
\end{eqnarray}
where $\theta_{ij},\;i,j=1,2,3$ are real angles and the phase $\delta_{CP}$ 
lies in the range  $0\leq\delta_{CP}\leq 2\pi$. In order to get nonzero
$CP$ violation, the quark mass matrices must have non-degenerate elements and
also $\theta_{ij}\neq0,\pi, \;i,j=1,2,3$, while  $\delta_{CP}\neq 0,\pi$.

According to the analysis by Shaposhnikov \cite{Shaposhnikov:1991cu}, the 
strength of the $CP$ violation in the standard electroweak model even at high
temperatures appears to be too small, so it is unlikely that it can 
explain the observed baryon asymmetry. This conclusion results from arguing 
that the only natural scale for the baryogenesis problem is the temperature
of the electroweak phase transition, which as we show in the Section~\ref{Sect:EWPT}
is of an order $T\sim 100\,\textrm{GeV}$. At this temperature the Yukawa interaction can 
be treated as a perturbation, because the quark masses are small compared with
the temperature. The baryon asymmetry as defined in Section~\ref{Sect:Baryogenesis-Problem} is
a dimensionless number, so to get an estimation of the asymmetry, the quantity
given by Eq.~(\ref{Eq:Delta-CP}) should be divided by something with dimensions of 
$(mass)^{12}$ and a natural choice is the temperature itself \cite{Farrar:hn}. Hence, 
\begin{equation}
\frac{n_B}{s}\leq\frac{\Delta_{CP}}{N_{eff}T^{12}}\sim 10^{-20}~,
\end{equation}
where $N_{eff}$ is the number of degrees of freedom  in the standard model.

The same author suggested in subsequent work 
\cite{Shaposhnikov:tw,Shaposhnikov:1987pf} a possible way out of this
problem. He assumed that accumulation of small $CP$ violating effects
during the expansion of the universe could enhance the total $CP$
violation during the electroweak phase transition, but this result is not
acceptable in general. Recent investigations show that the most 
promising models are models with enhanced the scalar sector and in particular
the models with two scalar Higgs doublets \cite{Cohen:1991iu}.

%% file: chap4.tex
\section{\label{Sect:Veff}The Effective Potential}

\subsection{Introduction}

At the very early stages of its evolution  
the universe is filled with matter at very high energies
and densities, so matter should be described in terms of quantum fields.
In order to investigate how the electroweak phase transition
could have occurred at the beginning of the universe 
we need  to work in the framework of quantum field theory.
The basic tool for the investigation of the nature of the electroweak phase
transition is the  effective potential, the quantity which has the meaning
of the potential energy density of the system under consideration.

The discussion on spontaneous symmetry breaking in Section~\ref{Sect:SSB} was 
purely classical. The particle spectrum was determined by minimization of 
the classical potential $V_{0}(\phi)$ as it appears into the Lagrangian  
and describes the potential energy density of a constant
scalar field. The effective potential has also the meaning of a potential
energy density. A quantum field theory involves virtual 
particles, which affect the field energy density through emission and 
reabsorption processes. This generalization of the classical potential  
to include  the 
quantum corrections is known as the effective potential. Minimization of 
the effective potential gives the field configuration with 
the minimal energy, the vacuum of the theory.  

Proceeding further, analysis of matter behaviour at non zero temperatures 
involves thermal fluctuations of the fields that one should take into 
account. Thus, a generalization of the effective potential at finite 
temperature is needed, for the inclusion of temperature dependent quantum 
effects. As it will be clear in what follows from the mathematical definition 
the effective action has the meaning of the free energy of the quantum system 
under consideration. The finite temperature effective potential $V(\phi, T)$ 
as Linde states \cite{Linde:1990} at its extrema coincides with the free energy 
density.

The effective potential has been studied extensively in the literature. An 
elegant discussion on 
the physical meaning of the effective potential and its calculation
is explored by Coleman \cite{Coleman:1988}.
A detailed analysis of the theory of the effective potential at zero and 
finite temperature with 
applications on cosmological models
is given by Brandenberger \cite{Brandenberger:cz}. 
The electroweak Higgs potential for the standard model and its extensions
has been investigated by Sher \cite{Sher:1988mj}. Early 
discussions on the subject are those of Coleman and Weinberg 
\cite{Coleman:jx} involving
calculation of the effective potential for a general gauge theory at zero
temperature. Generalization of the effective potential at finite temperature 
is given by Dolan and Jackiw \cite{Dolan:qd} and Linde 
\cite{Linde:1990,Linde:px}. In what follows we give a formal discussion on  the notion of 
the effective potential as it appears in the framework of quantum field 
theory.

The outline of this section is as follows: In Section~\ref{Sect:Veff-0} we discuss the
properties of the effective potential at zero temperature and give 
the basic steps for its 
calculation in  the framework of quantum field theory, using the path
integral formalism. For this analysis we restrict ourselves to the one loop 
radiative corrections. In Section~\ref{Sect:Veff-T} we generalize the idea of the
effective potential at finite temperature. This generalization appears in a form of integral terms which 
we add to the zero temperature effective potential. At high temperatures 
these integrals can be approximated expanding them in
a series. The high 
temperature approximation of 
the fermion and boson contribution to the effective potential is compared 
with the exact calculation of the one loop
radiative correction integrals using numerical methods.

\subsection{\label{Sect:Veff-0}Zero Temperature Effects}     

\subsubsection{Path Integral Formalism}

The effective potential can be calculated in an elegant way using the path 
integral formalism and the notion of  generating functionals.   
Consider the  case of a real scalar field and suppose that an external 
c--number source  $J(x)$, a function of space and time, is added into the 
Lagrangian coupled linearly to the field $\phi$. Then the transition 
amplitude from the vacuum state in the far past to the vacuum state in the 
far future is defined as
\begin{equation}
Z[J]=\int\mathcal{ D}\phi \exp(i\int d^{4} x[\mathcal{ L}(\phi)+J(x)\phi(x)])~.
\end{equation} 
This amplitude can be expanded  in a functional Taylor series in 
terms of the source $J(x)$, 
with coefficients  $G_{u}^{n}(x_{1},x_{2},\ldots,x_{n})$  the  Green 
functions of the theory. These Green functions can be found by functional 
differentiation  of $Z[J]$ with respect to $J(x)$, at $J(x)=0$.  

One can introduce  
the  generating functional of the connected Green functions $W[J]$, which 
is related to $Z[J]$ by 
\begin{equation}
Z[J]=\exp(iW[J])~.
\end{equation} 
Expanding this functional  in powers of $J(x)$, the coefficients
$G_{c}^{n}(x_{1},x_{2},\ldots,x_{n})$ 
are the connected Green's functions.  
The connected Green's functions $G^{(n)}_{c}$ are given by the functional 
derivative of $W[J]$ with respect to $J(x)$, at $J(x)=0$.  

At this point we have to introduce a quantity which  is referred as the 
classical field $\phi_{c}(x)$. This is a functional of the source $J(x)$
and is defined as the vacuum expectation value of the field operator
in the presence of the source as
\begin{equation}
\phi_{c}(x)=\frac{\delta W[J]}{\delta J(x)}~.
\end{equation}
In the absence of the source, the classical field $\phi_{c}(x)$
is just the vacuum expectation value of the field operator.

The generating functional of the 1PI (one particle irreducible) Green 
functions is called the effective action and is defined  by the 
functional Legendre transform 
\begin{equation}
\Gamma[\phi_{c}]=W[J]-\int d^{4}xJ(x)\phi_{c}(x)~. 
\end{equation} 
It is called the  effective action, because it is a functional of the classical
field $\phi_{c}$ and hence akin to the classical action $S[\phi]$. 
From the definition of the effective action, it follows that the source
can be obtained in the form of a functional derivative as
\begin{equation}
\frac{\delta\Gamma[\phi_{c}]}{\delta\phi_{c}(x)}=-J(x)~.
\end{equation}
The effective action can be expanded in powers of $\phi_{c}$, or alternatively
one can expand about a constant value of the  field $\phi_{c}$. This is the
same as expanding in powers of the derivatives of $\phi_{c}$. Thus in position 
space it is an expansion of the type
\begin{equation}
\Gamma[\phi_{c}]=\int d^{4}x[-V(\phi_{c})+\frac{1}{2} (\partial_{\mu}\phi_{c}
)^{2}Z(\phi_{c})+\ldots]~,
\end{equation}
where $V(\phi_{c})$ and  $Z(\phi_{c})$ are ordinary functions of $\phi_{c}$ , 
not functionals, and the function $V(\phi_{c})$ is known as  the effective 
potential. In the case of a classical
field $\phi_{c}$ which is constant in space and time and 
in the absence of the source $J$,  $\phi_{c}$ has the significance
of the vacuum expectation value (VEV) of the field operator
\begin{equation}
\frac{dV(\phi_{c})}{d\phi_{c}}=0~.
\end{equation}
This last equation allows one to interpret the VEV of the field 
operator as the stationary point of the effective potential, which can
be obtained by solving the above equation. Of course, in the case when
the field is to vary in space or time,  we have to solve the more
general equation $\delta \Gamma/\delta\phi_{c}=0$. Thus the VEV of
the field operator, taking into account quantum corrections, can be found 
by minimizing the effective potential. If the effective potential has several
local minima, it is the absolute minimum that corresponds to the 
true ground state.
  
It is useful to see that the effective potential can be obtained as the 
infinite sum of the 1PI graphs with vanishing external 
momenta \cite{Coleman:jx}, although we will not make use of this 
method in order to evaluate that 
in the case of the scalar field.  One can 
expand the effective action functionally in terms of $\phi_{c}$ as 
\begin{eqnarray}
\Gamma [\phi_{c}]&=&\sum_{n=1}^{\infty}\frac{1}{n!} \int
d^{4}x_{1}\ldots d^{4}x_{n}\;\phi_{c}(x_{1})\ldots\phi_{c}(x_{n})\nonumber \\
&&\times\Gamma^{(n)}(x_{1},\ldots,x_{n})~.
\end{eqnarray}
The coefficients $\Gamma^{(n)}$ are referred as the one--particle--irreducible 
(1PI)  Green functions.
Fourier transforming the 1PI Green functions to momentum space, the expression 
for the effective action takes the form
\begin{eqnarray}
\Gamma[\phi_{c}]&=&\sum_{n=1}^{\infty}\frac{1}{n!}\int 
d^{4}p_{1}\ldots d^{4}p_{n}\;\delta^{4}(p_{1}+\ldots +p_{n}) \nonumber \\
&&\times\Gamma^{(n)}(p_{1},\ldots, p_{n})\;\tilde{\phi_{c}}(p_{1})\ldots\tilde{\phi_{c}}(p_{n})~,
\end{eqnarray}
where $\tilde{\phi_{c}}(p_{i})$ is the Fourier transform of $\phi_{c}$.
Thus, comparing this last expression  with the momentum space expansion of 
the effective action above, we  find that the effective potential is given
as a sum of the 1PI graphs of the form
\begin{equation}
V(\phi_{c})=-\sum_{n=1}^{\infty}\frac{1}{n!}\Gamma^{(n)}(p_{i}=0)\;\phi^{n}_{c}~.
\end{equation}

\subsubsection{Scalar Loops}

In order to understand how the effective potential is calculated, 
we present here  the evaluation of the effective potential for a  
theory involving a real scalar field with quartic self interaction.
The Lagrangian governing this theory is given by
\begin{equation}
\mathcal{L}=\frac{1}{2}(\partial_{\mu}\phi)^{2}+\frac{1}{2}\mu^{2}\phi^{2}-
       \frac{1}{4}\lambda\phi^{4}~.
\end{equation}
The potential energy density of the real scalar field at classical level, 
which is commonly called the classical potential
and denoted by $V_{0}(\phi)$, is given by
\begin{equation}
V_{0}(\phi)=-\frac{1}{2}\mu^{2}\phi^{2}+\frac{1}{4}\lambda\phi^{4}~,
\end{equation}
where the field mass squared $\mu^{2}$ and the coupling constant $\lambda$
are positive. The effective potential which is usually denoted by $V(\phi)$
will result in as a sum of the classical potential, plus terms due to radiative
corrections \cite{Itzykson:rh,Bailin:wt,Ryder:wq,Ramond:pw,Rivers:hi}. 

As it was referred earlier the
diagrammatic method of summing the 1PI graphs can be used for the 
calculation of  the effective potential, however  we present here
an alternative method which is based on the saddle point evaluation of path 
integrals \cite{Ryder:wq,Ramond:pw}.
According to this method exponential integrals involving a function $f(x)$ 
which is stationary at some point 
$x_{0}$, can be solved by expanding $f(x)$ about this point. Then, omitting 
higher derivatives, the integral becomes a Gaussian and thus can be evaluated  
according to standard methods.

The starting point is the generating functional of the connected Green's
functions, where we have restored the Planck's constant so it takes the 
form 
\begin{equation}
\exp\bigg(\frac{i}{\hbar}W[J]\bigg)=\mathcal{N}\int\mathcal{D}\phi \exp\bigg(\frac{i}{\hbar}
S[\phi, J]\bigg)
\end{equation}
and the normalization constant $\cal N$ is chosen such as $W[0]=1$.
Although in natural units  the Planck constant is   
equal to unity, we restore it here in order to get an expansion in terms
of $\hbar$ and make clear that we calculate quantum corrections to the 
classical potential. The powers of $\hbar$ count the number of the closed
loops  in the loop expansion.                               
The action $S[\phi, J]$ in  presence of the source $J(x)$ is given by
\begin{equation}
S[\phi, J]=\int d^{4}x [\mathcal{ L}(\phi(x))+ J(x)\phi(x)]~.
\end{equation}
The saddle point is at $\phi=\phi_{0}$ and satisfies
\begin{equation}
\frac{\delta S[\phi,J]}{\delta \phi(x)}\bigg |_{\phi=\phi_{0}}=\hbar J(x)~,
\end{equation}
where $\phi$ is a function of $x$ and also a functional of $J$. In the case 
when the source $J$ tends to zero, $\phi_{0}(x)$ becomes a solution to the 
classical equations of motion. Expanding the action $S[\phi, J]$ about the 
stationary point at $\phi=\phi_{0}$ yields
\begin{eqnarray*}
S[\phi,J]& = & S[\phi_{0},J]+\int dx [\phi(x)-\phi_{0}] S'[\phi_{0},J] \\
         &   & \mbox{}+\frac{1}{2}\int dxdy[\phi(x)-\phi_{0}]
[\phi(y)-\phi_{0}]S''[\phi_{0},J]\nonumber~,
\end{eqnarray*}
plus higher order terms which we omit, since we are interested in  
calculating radiative corrections up to one loop only. In this last equation we have 
used the shorthand
\begin{eqnarray*}
S'[\phi_{0},J]=\frac{\delta S}
{\delta \phi(x)}\bigg |_{\phi=\phi_{0}}~,\\
S''[\phi_{0},J]=\frac{\delta^{2} S}{\delta\phi(x)\delta\phi(y)}\bigg |_
{\phi=\phi_{0}}~.
\end{eqnarray*}
We then  differentiate functionally  the action   
and insert the result  into the  equation above. The resulting integral
is a Gaussian one, so going over to Euclidean space we get finally the loop 
expansion for $W[J]$, ignoring
correction terms of order $\hbar^{2}$ and higher,
\begin{equation}
W[J]=S[\phi_{0}]+\hbar\int dx \phi_{0}(x)J(x)+\frac{i\hbar}{2} \textrm{Tr} \ln[\Box
+V_{0}^{\prime\prime}(\phi_{0})]~.
\end{equation}

Inserting this expression into the
defining equation of the effective action we get, by  putting the
source $J\rightarrow 0$,
\begin{equation}
\Gamma[\phi_{c}]=S[\phi_{c}]+\frac{i\hbar}{2} \textrm{Tr} \ln[\Box+V_{0}^{\prime\prime}
(\phi_{0})]~.
\end{equation}
For a field which is constant in space and time, through the expansion
of the effective action we find for the effective potential the expression
\begin{equation}
V(\phi)=V_{0}(\phi)+\frac{\hbar}{2}\int\frac{dk^{4}_{E}}{(2\pi)^{4}}\ln
[k^{2}_{E}+V_{0}^{\prime\prime}(\phi)]~,
\end{equation}
where in the final step we have used that the trace of an operator is the 
sum over its eigenvalues and we expressed the result on going over to 
Euclidean momentum space.

\subsubsection{Renormalization}

The above integral is divergent as  it happens in general when one
calculates radiative corrections  involving 
integrations over the internal momenta of the graphs.
For the theory to be renormalized, the divergences
must be absorbed, if possible, into the parameters of the theory.
In order to evaluate this integral we introduce a cut--off at some large
momentum $k^{2}=\Lambda^{2}$, so we obtain
\begin{eqnarray}
V_{1}(\phi)&=&V_{0}(\phi)+\frac{\Lambda^{2}}{32\pi^{2}}m^{2}(\phi) \nonumber \\
&&+\frac{m^{4}(\phi)}{64\pi^{2}}\bigg[\ln\bigg(\frac{m^{2}(\phi)}
{\Lambda^{2}}\bigg)-\frac{1}{2}\bigg]+V_{ct}(\phi)~,
\end{eqnarray}
where the field dependent squared mass of the scalar $m(\phi)$ is defined 
as the second derivative of the classical potential as
\begin{equation}
m^{2}(\phi)=\frac{d^{2}V_{0}(\phi)}{d\phi^{2}}=3\lambda\phi^{2}-\mu^{2}.
\end{equation}
To remove the cut--off dependence we have introduced a counterterm potential
which has the same structure as the original potential
\begin{equation}
V_{ct}(\phi)=\frac{1}{2}A\phi^{2}+\frac{1}{4}B\phi^{4}+C~.
\end{equation}
We can determine the coefficients  $A$ and $B$ by requiring that the position 
of the minimum of the effective potential
and the Higgs field mass remain in their classical values 
\cite{Linde:1990, Linde:px}, so
\begin{equation}
\frac{d V_{1}(\phi)}{d \phi} \bigg |_{\phi_{c}=\sigma}=0
\end{equation}
and the Higgs field mass $m^{2}_{H}=2\mu^{2}$ results as the second derivative
of the potential evaluated at $\phi_{c}=\sigma$,
\begin{equation}
\frac{d V^{2}_{1}(\phi)}{d \phi^{2}} \bigg |_{\phi_{c}=\sigma}=2\mu^{2}~.
\end{equation}                                                 
The final expression of the effective potential for the real scalar 
field, including radiative corrections up to one loop, takes the form
\begin{eqnarray}
V_{1}(\phi)&=&V_{0}(\phi)+\frac{m^{4}(\phi)}{64\pi^{2}}\ln\bigg(\frac{m^{2}(\phi)}
{m^{2}(\sigma)}\bigg)\nonumber \\
&&-\frac{1}{32\pi^{2}}m^{2}(\phi)m^{2}(\sigma)      
-\frac{3}{128\pi^{2}}m^{4}(\phi)~,
\end{eqnarray}
where $V_{0}(\phi)$ is the classical potential.
Other renormalization conditions can also be introduced 
\cite{Coleman:jx, Sher:1988mj} but  the physical results are meant 
to be the same.  

\subsubsection{Fermion and Boson Loops}
                               
The saddle point evaluation of the one loop corrections to the 
effective potential was an elegant way to obtain the effective potential in 
the case of the one scalar field. In order to proceed and obtain an 
expression of  the effective potential for the standard electroweak model we
need to know the one loop contributions of fermions and vector bosons.
The case of a general non-abelian gauge theory was the first discussed by 
Coleman and Weinberg \cite{Coleman:jx}. An analytic presentation
of the subject is also given by  Rivers
\cite{Rivers:hi} and Sher \cite{Sher:1988mj}. In this general  case it is more 
convenient to revert to the definition of 
the effective potential from the expansion of the effective action and
the first step is to extend the scalar sector including
more fields. 
A  different approach, based on the evaluation of Gaussian integrals as in the
case of the scalar field, is given in Bailin and Love \cite{Bailin:wt}.
 
According to Anderson and Hall \cite{Anderson:1991zb}, the  analysis  given in the 
above references can be summarized into that the unrenormalized one loop 
self--energy contribution of  virtual particles, adds to the classical
potential a term which in the case of fermions is given by
\begin{eqnarray}
\Delta V_{1f}(\phi)&=&-\frac{1}{64\pi^{2}}m^{4}_{f}(\phi)\ln\bigg(
\frac{m^{2}_{f}(\phi)}{\Lambda^{2}}\bigg) \nonumber \\
&&+\frac{1}{128\pi^{2}} m^{4}_{f}(\phi)   
-\frac{1}{32\pi^{2}}m^{4}_{f}(\phi)\Lambda^{2}
\end{eqnarray}
for each fermionic degree of freedom.
On the other hand the bosons contribution has a similar form, except the 
opposite sign, and is given by
\begin{eqnarray}
\Delta V_{1b}(\phi)&=& \frac{1}{64\pi^{2}}m^{4}_{b}(\phi)\ln\bigg(
\frac{m^{2}_{b}(\phi)}{\Lambda^{2}}\bigg)\nonumber \\
&&-\frac{1}{128\pi^{2}} m^{4}_{b}(\phi)   
-\frac{1}{32\pi^{2}}m^{2}_{b}(\phi)\Lambda^{2}~.
\end{eqnarray}
In both the above equations $\Lambda$ is a cut--off which has been introduced 
in order to evaluate the divergent integrals.

Summarizing the above discussion, the one loop effective potential in the 
case of a general gauge theory
involving fermion, scalar and boson fields can be written in the general
form
\begin{equation}
V_{1}(\phi)=V_{0}(\phi)+\Delta V_{1}(\phi)+V_{ct}(\phi)~,
\end{equation}
where $\Delta V_{1}(\phi)$ is the sum of fermions, scalars and bosons
one loop contributions to the effective potential and $V_{ct}(\phi)$ is  a 
counterterm potential which is used to remove the cut--off dependence as in
the scalar field case.
We adopt the same renormalization condition as before in order to retain 
the Higgs mass and the position of the minimum of the potential for
each degree of freedom to which the scalar couples, so for the bosons
we have a contribution of the form

\begin{eqnarray}
\Delta V_{1b}&=& \sum_{b}\frac{n_{b}}{64\pi^{2}} \bigg [ m^{4}_{b}(\phi)\ln\bigg(
\frac{m^{2}_{b}(\phi)}{m^{2}_{b}(\sigma)}\bigg)-\frac{3}{2} m^{4}_{b}(\phi) \nonumber \\  
&& +2 m^{2}_{b}(\phi)m^{2}_{b}(\sigma) \bigg ]~,
\end{eqnarray}
while for the fermions the contribution reads as
\begin{eqnarray}
\Delta V_{1f}&=&-\sum_{f}\frac {n_{f}}{64\pi^{2}}\bigg [ m^{4}_{f}(\phi)\ln\bigg(
\frac{m^{4}_{f}(\phi)}{m^{2}_{f}(\sigma)}\bigg)-\frac{3}{2} m^{4}_{f}(\phi)\nonumber \\     
&&+2 m^{2}_{f}(\phi)m^{2}_{f}(\sigma)\bigg ]~,
\end{eqnarray}
where $n_{b}$ and $n_{f}$ are the numbers of degrees of freedom associated
to spin, particle--antiparticle states  and internal symmetries (coloured 
quark states).

\subsection{\label{Sect:Veff-T}Finite Temperature Effects}

\subsubsection{Fields at Finite Temperature }

In order to investigate the electroweak phase transition in the next section,
we need a 
generalization of the effective potential at finite temperatures. We present 
here
how the notion of the effective potential can be generalized at finite 
temperature, following the ideas of Dolan and Jackiw \cite{Dolan:qd}, 
Linde \cite{Linde:1990,Linde:px} and Sher \cite{Sher:1988mj}.

At finite temperature a field theory is equivalent to an ensemble of finite 
temperature Green functions. The average value of an operator at nonzero 
temperature is defined by the Gibbs average as
\begin{equation}
\mathcal{O}(x_{1},x_{2},\ldots,x_{n})=\frac{\textrm{Tr} e^{-\beta H}\mathcal{O}(x_{1},x_{2}
\ldots,x_{n})}{\textrm{Tr}e^{-\beta H}}~,
\end{equation}
where $H$ is the Hamiltonian of the system under consideration and
$\beta=1/k_{B}T=1/T$, since for the sake of simplicity  
the Boltzmann's constant $k_{B}$ can be taken equal to unity. Then, according 
to the proof given by Sher \cite{Sher:1988mj}, Green's functions at 
non zero temperature obey the same equations as those at zero temperature, but 
under  different  boundary conditions. The finite temperature Green's 
functions concerning bosons  are periodic in Euclidean time, $\tau\equiv it$, 
with a  period $\beta=1/T$, instead of having the usual boundary conditions
$t=\pm\infty$. On the other hand  fermionic finite temperature Green's
functions obey antiperiodic boundary conditions with the same
period $\beta$. 

In  analogy with the previous section 
the finite temperature effective potential can be calculated by using
similar methods as in zero temperature field theory and the path 
integral formalism. A finite temperature effective action 
$\Gamma^{\beta}[\phi_{c}]$
is defined by analogy to that at zero temperature, since  the vacuum
expectation value of the classical field at zero temperature now corresponds 
to a thermodynamic average. The finite temperature
effective potential may be defined by an expansion of the effective action
analogous to that of the previous section and its calculation proceeds through 
the evaluation of Gaussian path integrals at Euclidean space--time. 
The only difference in these calculations is that when  one calculates 
integrals involving boson fields  all the boson momenta $k_{0}$ should be 
replaced by the Matsubara frequencies for bosons $2\pi nT$ with $n$ an integer. 
On the other hand, since fermionic fields are 
antiperiodic, the  corresponding fermion momentum $k_{0}$ must be replaced
by $(2n+1)\pi T$. So finally, instead of integrating over $k_{0}$, one has 
to sum over $n$. We  show this calculation in some detail and give 
relevant references in the following sections.

\subsubsection{Scalar Fields at Finite Temperature}

Consider first the case of a scalar field. We have to expand the Lagrangian
around  a constant field $\phi_{c}$. The field dependent mass squared, which
is called the effective mass, is given by
\begin{equation}
m^{2}(\phi_{c})=3\lambda\phi_{c}^{2}-\mu^{2}~.
\end{equation}
The zero loop effective potential, the so--called classical potential, is 
temperature independent and is given by
\begin{equation}
V_{0}(\phi_{c})=-\frac{1}{2}\mu^{2}\phi_{c}^{2}+\frac{1}{4}\lambda\phi^{4}_{c}~.      
\end{equation} 
We have already calculated the one loop approximation  to the effective 
potential at zero temperature as 
\begin{equation}
V^{0}_{1s}(\phi_{c})=\frac{1}{2} \int \frac{d^{4}k}{(2\pi)^{4}}\ln[ k^{2}
+m^{2}(\phi_{c})]~.
\end{equation}
At finite temperature the above expression, as we have stated earlier, 
one has to replace the scalar boson momenta $k_{0}$ by the Matsubara 
frequency  $2\pi n T$, so it becomes
\begin{equation}
V^{T}_{1s}(\phi_{c})=\frac{1}{2}T\sum_{n=-\infty}^{\infty}\int\frac{d^3\textbf{k}}
{(2\pi)^{3}}\ln[\textbf{k}^{2}+(2\pi  nT)^{2}+m^{2}(\phi_{c})]~.
\end{equation} 
The sum over $n$ diverges, so in order to evaluate this integral one can
follow the procedure given by Dolan and Jackiw \cite{Dolan:qd}. The result splits 
in two parts, one temperature independent part
\begin{equation}
V^{0}_{1s}(\phi_{c})=\int\frac{d^{3}\textbf{k}}{(2\pi)^{3}}
\frac{\sqrt{\textbf{k}^{2}+m^{2}(\phi_{c})}}{2}~,
\end{equation}
which is equivalent to the one loop effective potential at zero temperature
which  we have calculated  in the previous section, and a temperature dependent
part
\begin{equation}
V^{T}_{1s}(\phi, T)=\frac{T^{4}}{2\pi^{2}}\int_{0}^{\infty}dx\,x^{2}\ln[1-
\exp(-A(x,m,T))]~.   
\end{equation}
In this last equation  we have introduced the shorthand expression 
$A(x,m,T)=\sqrt{x^{2}+m^{2}(\phi_{c})/T^{2}}$~.
These final steps are explained in some detail in the next section where we
give the basic steps for the calculation of the finite temperature effective 
potential for a general gauge theory. 

Thus summarizing the above 
discussion, the finite temperature effective potential for a real scalar 
field including
the one loop radiative corrections can be written as
\begin {equation}
V_{1s}(\phi, T)=V_{0}(\phi)+V_{1s}^{0}(\phi)+V_{1s}^{T}(\phi, T)~.
\end{equation}
The finite temperature contribution vanishes as it should at zero temperature
$T=0$, when the mass squared $m^{2}(\phi_{c})$ is positive. But as it was 
stressed by Dolan and Jackiw \cite{Dolan:qd},
if one  is
to use this full expression of the one loop potential, there are some serious
difficulties for small values of the field, $\phi^{2}<\mu^{2}/3\lambda$, since 
the mass squared  of the Higgs scalar becomes negative leading
to an unacceptable complex one loop effective potential.

Fortunately in the case of the minimal standard model, which we discuss 
in the next section, the scalar loops can be safely  ignored, since they 
are negligible as compared to
fermion and vector boson contributions, simply because there is a large
degeneracy factor associated to the latter, while there is only one Higgs
scalar.

\subsubsection{Fermions--Bosons at Finite Temperature} 

The above discussion can be applied to a general gauge theory and as we 
stated earlier
there is an analogy between the computation of the effective potential at
finite temperature to that at zero temperature.  
We have to expand the scalar fields around their expectation values, which 
are now 
thermal averages, and isolate the terms in the Lagrangian which are quadratic
in all the fields. Then the one loop contribution to the effective action
is obtained as a Gaussian path integral. Suppose we deal with a general gauge
theory involving fermions, scalars, gauge bosons and Fadeev--Popov ghost fields.
We  insert the  quadratic Lagrangian into  the expression 
of the effective action and evaluating the resulting Gaussian integrals, the 
finite temperature effective potential results as a sum of traces of
the form \cite{Bailin:wt}
\begin{equation}
-\int_{0}^{\beta}d\tau\int d^{3}x V^{T}_{1}(\phi_{c})=
-\frac{1}{2}\textrm{Tr}\ln\textbf{A}-\frac{1}{2}\textrm{Tr}\ln\textbf{B}+
\textrm{Tr}\ln\textbf{C}~.
\end{equation}
The scalar contribution is given by the first trace
\begin{equation}
\textrm{Tr}\ln\textbf{A}=\int d^{3}x\sum_{i}^{\textrm{scalar}}\int\frac{d^3\textbf{k}}{(2\pi)^
{3}}\sum_{n}\ln[\omega^{2}_{n}+\textbf{k}^{2}+(m^{2}_{s})_{i}]~,
\end{equation}
where $m_{s}^{2}=m_{s}^{2}(\phi_{c})$ are the 
eigenvalues of the scalar mass matrix.            
The contribution of the vector bosons has the form
\begin{equation}
\textrm{Tr}\ln\textbf{B}=\int d^{3}x\sum^{\textrm{boson}}_{a}\int\frac{d^3\textbf{k}}{(2\pi)^
{3}}\sum_{n}3\ln[\omega_{n}^{2}+\textbf{k}^{2}+(m_{b}^{2})_{a}]~,
\end{equation}
with $m_{b}^{2}=m_{b}^{2}(\phi_{c})$ being the eigenvalues of the vector 
boson mass matrix. 
In a similar way the fermion term appears as
\begin{equation}
\textrm{Tr}\ln\textbf{C}=2\int d^{3}x\sum_{r}^{\textrm{fermion}}\int\frac{d^3\textbf{k}}{(2
\pi)^{3}}\sum_{n}\ln[\omega_{n}^{2}+\textbf{k}^{2}+(m^{2}_{f})_{r}]~.
\end{equation}
These can be evaluated by using the Matsubara frequency sums  
which in the case of bosons  is given by
\begin{equation}
\sum_{n}\ln(\omega_{n}^{2}+x^{2})=\beta x+2\ln(1-e^{-\beta x})
\end{equation}
plus an $x$--independent constant. In the 
fermion case the relevant
formula has a similar form apart from a minus sign, 
\begin{equation}
\sum_{n}\ln(\omega_{n}^{2}+x^{2})=\beta x+2\ln(1+e^{-\beta x})~,
\end{equation}
where  for the antiperiodic fermions $\omega_{n}=(2n+1)\pi T$.

After summing, the scalar contribution takes the form
\begin{eqnarray}
\Delta V_{1s}(\phi_{c}, T)&=&\frac{1}{2}\int\frac{d^3\textbf{k}}{(2\pi)^{3}}
\sum_{i}\bigg [(E^2_s)_i \nonumber \\
&&+\frac{2}{\beta}\ln \big [1-\exp[-\beta (E^2_s)_i]\big ] \bigg ]~,  
\end{eqnarray}
the vector bosons contribute the following term
\begin{eqnarray}
\Delta V_{1b}(\phi_{c}, T)&=&\frac{1}{2}\int\frac{d^3\textbf{k}}
{(2\pi)^{3}}\sum_{a}\bigg [3 (E^2_b)_a \nonumber \\
&&+\frac{6}{\beta}\ln \big [1-\exp[-\beta (E^2_b)_a]\big ] \bigg ]~, 
\end{eqnarray}
and the contribution of the fermions reads as  
\begin{eqnarray}
\Delta V_{1f}(\phi_{c}, T)&=&-2\int\frac{d^3\textbf{k}}{(2\pi)^{3}}
\sum_{r}\bigg [(E^2_f)_r \nonumber \\
&&+\frac{2}{\beta}\ln \big [1+\exp[-\beta (E^2_f)_r]\big ] \bigg ]~,
\end{eqnarray}
where we have used that $E^2_a=\sqrt{\textbf{\textrm{k}}^2+ m_a^2}$ for $a=s, b, f$
for scalars, boson and fermions.
All  the above equations  can be separated into
a part $\Delta V_{1i}^{0}(\phi), i=s, b, f$ which is temperature independent 
and a temperature
dependent part $\Delta V_{1i}^{T}(\phi)$ for each of the above field 
contributions.

By using the fact that \cite{Dolan:qd,Bailin:wt} 
\begin{equation}
\int\frac{d^3\textbf{k}}{(2\pi)^{3}}\sqrt{\textbf{k}^{2}+R}    
=\int\frac{d^{4}k}{(2\pi)^{4}}\ln(k_{0}^{2}+\textbf{k}^{2}+R)
\end{equation}
plus a constant independent of $R$ and evaluating the integrals in the 
zero temperature part, we get the one loop  quantum corrections which we 
have already calculated in the previous section.
By setting $x\equiv |\textbf{k}|/T$ in the temperature dependent part
of the above equations, the scalar field contribution to the effective 
potential at finite temperature takes  the form
\begin{eqnarray*}
\Delta V^{T}_{1s}(\phi, T)&=&\frac{T^{4}}{2\pi^{2}}\int_{0}^{\infty}
dx\;x^{2} \nonumber \\
&&\times\sum_{i}\ln \big [ 1-\exp(-\sqrt{x^{2}+(m_{s}^{2})_i/T^{2}}) \big ]~, 
\end{eqnarray*}    
the boson term is
\begin{eqnarray*}
\Delta V^{T}_{1b}(\phi, T)&=&3\frac{T^{4}}{2\pi^{2}}\int_{0}^{\infty}
dx\;x^{2}\nonumber \\
&&\times\sum_{a}\ln \big [ 1-\exp(-\sqrt{x^{2}+(m_{b}^{2})_a/T^{2}}) \big ]~,
\end{eqnarray*}    
and the fermion term is
\begin{eqnarray*}
\Delta V^{T}_{1f}(\phi, T)&=&-4\frac{T^{4}}{2\pi^{2}}\int_{0}^{\infty}
dx\;x^{2} \nonumber \\
&&\times\sum_{r}\ln \big [1+\exp(-\sqrt{x^{2}+(m_{b}^{2})_r/T^{2}}) \big ]~, 
\end{eqnarray*}    
where  $m_{i}=m_{i}(\phi_{c}), i=s, f, b$ are the scalar, fermion and boson
mass eigenvalues.  

It is easy to observe the similarity in the formulas between the vector bosons
and scalars. The only difference is the factor 3 which appears in the boson
case. This factor expresses the one longitudinal  and the two transverse
degrees of freedom of a massive vector boson. On the other hand the factor 
4 in the formula for fermions corresponds to the two fermionic spin states
times the two particle antiparticle states.
On summarizing the above discussion we can say that finite temperature
effects for bosons takes the form
\begin{eqnarray*}
\Delta V^{T}_{1b}(\phi, T)&=&\sum_{b}\frac{n_{b}T^{4}}{2\pi^{2}}\int_{0}^{\infty}
dx\;x^{2}\nonumber \\
&&\times\ln[1-\exp(-\sqrt{x^{2}+m_{b}^{2}(\phi)/T^{2}})]~. 
\end{eqnarray*}    
The fermion term has a similar form apart from the positive sign in 
the argument  of the logarithm and that their contribution has an overall
minus sign, so
\begin{eqnarray*}
\Delta V^{T}_{1f}(\phi, T)&=&-\sum_{f}\frac{n_{f}T^{4}}{2\pi^{2}}\int_{0}^{\infty}
dx\;x^{2} \nonumber \\
&&\times\ln[1+\exp(-\sqrt{x^{2}+m_{f}^{2}(\phi)/T^{2}})]~. 
\end{eqnarray*}    
As it was referred above for massive vector bosons the relevant factor which
corresponds to the particle helicity states is $n_{b}=3$, since for fermions 
$n_{f}=4$. For coloured fermions, as it is the case of quarks, $n_{f}=12$.   

\subsubsection{\label{Sect:Veff-HT}High Temperature Approximation}

It is sometimes convenient to approximate the above integrals by using the 
high temperature expansion. When the temperature is high enough
as compared to the particle masses the above integrals can be 
approximated by expanding them  in a series in powers of $m(\phi)/T$. This 
expansion has been extensively used in the literature 
in  investigations of the effective potential.
The expansion of the integrals in the high temperature limit can be done
by using the Riemann Zeta function \cite{Dolan:qd, Kapusta:tk}.
According to the analysis given in the above references, the integral in 
the fermion 
case can be expanded as
\begin{eqnarray*}
\Delta V_{1f}^{HT}(\phi, T)&=&\sum _{f}n_{f}\bigg [-\frac{7\pi^{2}T^{4}}{720}
+\frac{m^{2}_{f}T^{2}}{48} \nonumber \\
&&+\frac{m_{f}^{4}}{64\pi^{2}}\ln\bigg (\frac{m_{f}^{2}}{c_{f}
T^{2}}\bigg )+O\bigg (\frac{m_{f}^{6}}{T^{2}}\bigg)\bigg ]~,
\end{eqnarray*}
where the constant $c_{f}$ is defined as $\ln c_{f}=\frac{3}{2}+2
\ln\pi-2\gamma\approx 2.64 $ and the Euler constant is 
$\gamma\approx 0.577$. The particle masses which appear in the above expression 
are field dependent, so $m_{f}=m_{f}(\phi)$. 
                      
Our aim was to verify  the reliability of the high temperature expansion,
so we have calculated the integrals given in the previous section 
numerically. For this calculation we have used a numerical code based on
Simpson's rule. 

\begin{figure}
\includegraphics[scale=0.46]{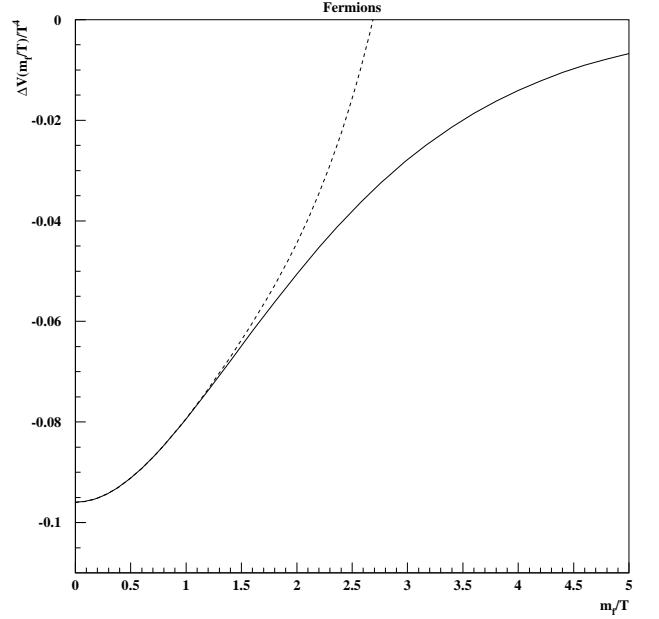} 
\caption{\label{Fig:41}A fermion's contribution to the effective potential as a function 
of $m_{f}(\phi)/T$. The solid line represents the exact calculation, while
the dashed line corresponds to the high temperature expansion.}
\end{figure}
In order to
compare  a fermion's contribution to the effective potential by using        
the high temperature approximation with the exact calculation of the 
integrals,  the contribution of a fermionic degree
of freedom to the free energy density (the effective potential) as a function
of $m_{f}/T$ is given in Fig.~\ref{Fig:41}.
As we can observe in this figure, for $m_{f}/T<1.6$, the high temperature 
approximation is in 
good agreement (better than 5\%) with the
exact calculation of the effective potential. 

\begin{figure}
\includegraphics[scale=0.46]{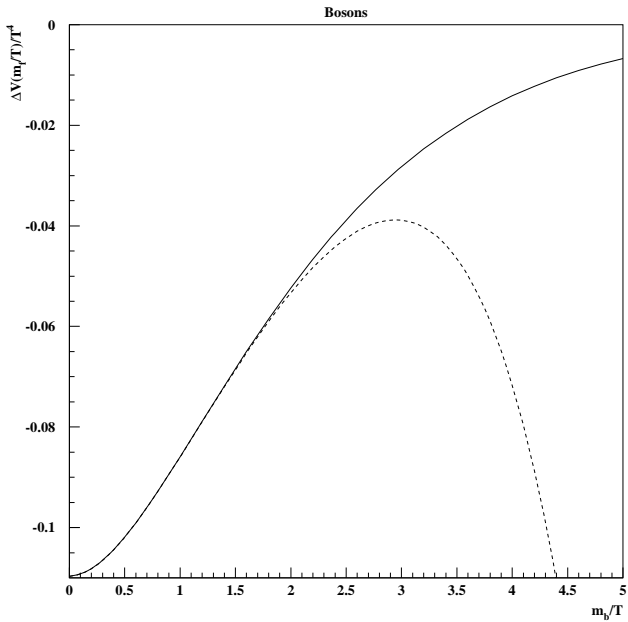} 
\caption{\label{Fig:42} A boson's contribution to the effective 
potential as a function 
of $m_{b}(\phi)/T$. The solid line represents the exact calculation, while
the dashed line corresponds to the high temperature 
expansion.}
\end{figure}

In the boson case the expansion of the integral reads 
\begin{eqnarray*}
\Delta V_{1b}^{HT}(\phi, T)&=&\sum _{b}n_{b}\bigg [-\frac{\pi^{2}T^{4}}{90}
+\frac{m^{2}_{b}T^{2}}{24} \nonumber \\
&&-\frac{m^{3}_{b}T}{12\pi}
-\frac{m_{b}^{4}}{64\pi^{2}}\ln\bigg(\frac{m_{b}^{2}}{c_{b}T^{2}}\bigg)
+O\bigg(\frac{m_{f}^{6}}{T^{2}}\bigg)\bigg]
\end{eqnarray*}
where the constant $c_{b}$ is defined as $\ln c_{b}=\frac{3}{2}+2
\ln4\pi-2\gamma\approx 5.41 $. 
As in the fermion case the boson masses are field
dependent, so $m_{b}=m_{b}(\phi)$. The fermion and boson expressions
look similar, but one can observe that there are no
cubic terms in the formula for fermions since there cannot be 
modes of zero Matsubara frequency. This last observation turns to be very 
crucial in the standard model case which we investigate in the next section.    
A boson's contribution to the effective potential is given in Fig.~\ref{Fig:42} and
from this picture it is clear that the high temperature expansion of 
the integrals is consistent with the exact calculation  to better than 5\% 
for $m_{b}/T<2.2$. The previous graphs confirm similar results  obtained by Anderson
and Hall \cite{Anderson:1991zb}. 

%% file: chap5.tex
\section{\label{Sect:EWPT}Electroweak Phase Transition}

\subsection{Introduction}

The third Sakharov condition, that the universe must be out of 
equilibrium, demands that the electroweak phase transition should be of 
the first order, if baryogenesis is to happen at this transition.
Although the order of the transition is not yet completely established and the
literature contains contradictory claims, most of the recent investigations
suggest that it is of the first order.
The basic tool for the investigation of the electroweak phase transition 
is the finite temperature effective potential. Our aim is to explore how the phase 
transition at the electroweak scale could have taken place and what is the
order of the transition when the one loop approximation is used in the  
calculation of the effective potential.

In the next Section~\ref{Sect:SM-Veff} we use the analysis given 
in Section~\ref{Sect:Veff} in order to calculate the      
one loop corrections to the effective potential of the standard model 
and obtain an  expression of the effective 
potential in the high temperature approximation. In Section~\ref{Sect:SMPT} we 
investigate the nature the electroweak phase transition through the
analysis of the evolution of the effective potential, while in Section~\ref{Sect:Bubble-Nucleation} 
we outline how the transition is dynamically achieved. 

\subsection{\label{Sect:SM-Veff}The Standard Model Effective Potential}

\subsubsection{Zero Temperature}

The calculations concerning the contributions to the effective potential of 
fermions and bosons  given 
in Section~\ref{Sect:Veff} can be applied in the case of the minimal standard
model of the electroweak interactions. The effective potential at zero 
temperature will appear as a sum of the classical potential $V_{0}(\phi)$
plus terms due to one loop radiative corrections of fermions
and bosons which are present at the standard electroweak model and where we
will assume that the contribution of the Higgs scalar is negligible.

Only the heaviest particles give a significant 
contribution to the effective potential so we have to include
the three gauge bosons, the two charged $W$ and the neutral $Z$, since
the only heavy fermion is the top quark. The top's mass is not known 
experimentally but there is  an experimental lower limit which is about 
$130\,\textrm{GeV}$  \cite{Abachi:1994je}.  By using the  results of Section~\ref{Sect:Veff-0}
the contribution of the two charged vector bosons 
takes the form
\begin{equation}
\Delta V_{1W}(\phi)=2\frac{3}{64\pi^{2}\sigma^{4}}m_{W}^{4}(\sigma)\left [
\ln\bigg(\frac{\phi^{2}}{\sigma^{2}}\bigg) -\frac{3}{2}\phi^{4}+2\sigma^{2}\phi^{2}
\right]
\end{equation}
where the factor 2 stands for the two charged particles and 3 corresponds 
to the polarization degrees of freedom. In this equation  we have expressed 
the  Ws'
contribution as a function of the field and the masses of the particles are
those at tree level given by   $m_{W}^{2}(\sigma)=1/4g^{2}\sigma^{2}$.       
In a similar way the contribution of the one neutral $Z$ boson is
\begin{equation}
\Delta V_{1Z}(\phi)=\frac{3}{64\pi^{2}\sigma^{4}}m_{Z}^{4}(\sigma)\left [
\ln\bigg(\frac{\phi^{2}}{\sigma^{2}}\bigg) -\frac{3}{2}\phi^{4}+2\sigma^{2}\phi^{2}
\right]
\end{equation}
and the $Z$ mass at the minimum of the potential is 
$m_{Z}^{2}(\sigma)=1/4(g^{2}+g^{\prime^{2}})^{2}\sigma^{2}$.
The top quark contributes the following term
\begin{equation}
\Delta V_{1t}(\phi)=-4\frac{3}{64\pi^{2}\sigma^{4}}m_{t}^{4}(\sigma)\left [
\ln\bigg(\frac{\phi^{2}}{\sigma^{2}}\bigg) -\frac{3}{2}\phi^{4}+2\sigma^{2}\phi^{2}
\right]
\end{equation}
with a  tree level mass $m_{t}^{2}(\sigma)=1/2g^{2}_{t}\sigma^{2}$, where 
$g_{t}$ is the top quark Yukawa coupling and the overall factor 12
corresponds to the degrees of freedom of the coloured top quark.

Adding  the above terms  the final expression of the effective potential for 
the standard  electroweak model including quantum corrections up to one loop
can be written  as
\begin{equation}
\Delta V_{1}^{0}(\phi)=V_{0}(\phi)+\Delta V_{1W}(\phi)+\Delta V_{1Z}(\phi)     
+\Delta V_{1t}(\phi)
\end{equation}
and in terms of the order parameter $\phi$
\begin{equation}
\Delta V^{0}_{1}(\phi)=-\frac{1}{2}\mu^{2}\phi^{2}+\frac{1}{4}\lambda\phi^{4}
+2B\sigma^{2}\phi^{2}-\frac{3}{2}B\phi^{4}+B\phi^{4}\ln\frac{\phi^{2}}
{\sigma^{2}}
\end{equation}
where  $B$ is defined as
\begin{equation}
B=\frac{3}{64\pi^{2}\sigma^{4}}(2m_{W}^{4}+m_{Z}^{4}-4m_{t}^{4})
\end{equation}
and $\sigma=246\,\textrm{GeV}$ is the value of the scalar field at the minimum of the
one loop potential $V^{0}_{1}(\phi)$ at zero temperature. The Higgs field 
coupling constant is defined as
$\lambda=\mu^{2}/\sigma^{2}$ and the Higgs boson mass is given by
$m_{H}^{2}=2\mu^{2}$. 
The masses of the particles which appear into the above expression for $B$ 
are the tree level masses, $m_{i}=m_{i}(\sigma), i=W, Z, t$.
Similar expressions up to a change of variables appear also in references 
\cite{Carrington:1991hz,Espinosa:1992gq,Anderson:1991zb,Dine:1991ck}

\subsubsection{Finite Temperature}

At finite temperature we use the results of Section~\ref{Sect:Veff-T} and we
add to the potential  a term which includes the temperature induced
effects. As we have discussed already only the heaviest particles give a 
significant contribution so we 
have to include only the two vector bosons and the top quark. As we stated
earlier the scalar contribution can be ignored. We can write  the 
finite temperature term in a compact form as
\begin{equation}
\Delta V_{1}^{T}(\phi, T)=\frac{T^{4}}{2\pi^{2}}[n_{W}J_{b}(y_{W})
+n_{Z}J_{b}(y_{Z})-n_{t}J_{f}(y_{t})]
\end{equation}
where $J_{b}(y)$ is the integral contribution of the bosons and is given by
\begin{equation}
J_{b}(y)=\int_{0}^{\infty}dx x^{2}\ln[1-\exp(-\sqrt{x^{2}+y^{2}})]
\end{equation}
and the fermion contribution, which differs only by the minus sign has the
form
\begin{equation}
J_{f}(y)=\int_{0}^{\infty}dx x^{2}\ln[1+\exp(-\sqrt{x^{2}+y^{2}})].
\end{equation}
The factors $n_{i}, i=W, Z, t$ are the number of the particle spin states,
so for the two charged $W$ vector bosons, times  three for the two transverse 
and the one longitudinal degrees of freedom we find  $n_{W}=6$, since the
relevant factor for the neutral $Z$ boson is $n_{Z}=3$.
For the top quark the relevant factor corresponds to the two spin states for 
fermions, times two particle-antiparticle states, times three
coloured quark states, so $n_{t}=12$. These factors justify our choice to
neglect the scalar loops, since  there  is only one scalar 
but nine vector contributions and twelve top quarks. The variable  $y$ is 
defined as the ratio of the field depended particle mass to the temperature
$y=m(\phi)/T$ and can also be written as a function of the field as
$y=m_{i}(\sigma)\phi/\sigma T$ where $m_{i}(\sigma)$ is the particle mass 
at the classical level. 

\subsubsection{High Temperature Expansion}

In the case when the temperature can be considered as high enough
compared to the particle masses, by using the results presented in 
Section~\ref{Sect:Veff-HT}, the above integrals can be expanded according to the
standard formulas. Summarizing the vector bosons and the top quark
contributions, the high temperature expansion of the integrals
can be written as
\begin{eqnarray*}
\Delta V^{HT}_{1}(\phi, T)&=&n_{f}\bigg[\frac{m_{t}^{2}(\phi)T^{2}}{48}
+\frac{m_{t}^{4}(\phi)}{64\pi^{2}}\ln\bigg(\frac{m_{t}^{2}(\phi)}{c_{f}T^{2}}\bigg)\bigg ]\nonumber\\
&&+\sum_{i=W, Z}n_{b}\bigg[\frac{m_{i}^{2}(\phi)
T^{2}}{24}-\frac{m_{i}^{3}(\phi)T}{12\pi}\nonumber\\
&&-\frac{m_{i}^{4}(\phi)}{64\pi^{2}}
\ln\bigg(\frac{m_{i}^{2}(\phi)}{c_{b}T^{2}}\bigg)\bigg ]
\end{eqnarray*}
where in the above expression we have omitted terms of order 
$O(m^{6}(\phi)/T^{2})$ or higher and terms which are independent of the field
$\phi$. Introducing the tree level masses as before, the above expression 
is written as a function of the field $\phi$ in the form
\begin{eqnarray*}
\Delta V^{HT}_{1}(\phi, T)&=& n_{f}\bigg [\frac{m_{t}^{2}\phi^{2}T^{2}}
{48\sigma^{2}}+\frac{m_{t}^{4}\phi^{4}}{64\pi^{2}\sigma^{4}}
\ln\bigg(\frac{m_{t}^{2}\phi^{2}}{c_{f}\sigma^{2}T^{2}}\bigg)\bigg ] \nonumber\\
&&+\sum_{i=W, Z}n_{b}\bigg [\frac{m_{i}^{2}\phi^{2}
T^{2}}{24\sigma^{2}}-\frac{m_{i}^{3}\phi^{3}T}{12\pi\sigma^{3}}\nonumber\\
&&-\frac{m_{i}^{4}\phi^{4}}{64\pi^{2}\sigma^{4}} 
\ln\bigg(\frac{m_{i}^{2}\phi^{2}}{c_{b}\sigma^{2}T^{2}}\bigg)\bigg ]
\end{eqnarray*}
    
In order to get a final expression for the full effective potential at the 
high temperature limit, we can 
write that as a sum of the zero temperature part plus the above high 
temperature expansion. As we can observe the field dependent logarithmic 
terms cancel
between the zero and high temperature parts and that there is no cubic 
term in the top quark contribution, due to that there cannot be 
modes of zero Matsubara frequency.  

The final expression of the finite temperature effective potential
at the high temperature limit appears as a sum of the effective potential
at zero temperature plus the above expansion of the integrals, so it has 
the compact form
\begin{equation}
\label{Eq:v(phi,T)}
V(\phi, T)=D(T^{2}-T_{0}^{2})\phi^{2}-ET\phi^{3}-\frac{1}{4}\lambda_{T}
\phi^{4}.
\end{equation}
The parameters of the above equation are defined as
\begin{equation}
D=\frac{1}{8 \sigma^{2}}(2m_{W}^{2}+m_{Z}^{2}+2m_{t}^{2}),
\end{equation}
\begin{equation}
E=\frac{1}{4\pi \sigma^{3}}(2m_{W}^{3}+m_{Z}^{3}),
\end{equation}
\begin{equation}
T_{0}^{2}=\frac{1}{2D}(\mu^{2}-4B\sigma^{2})
=\frac{1}{4D}(m_{H}^{2}-8B\sigma^{2}).
\end{equation}
The temperature dependent coupling constant $\lambda_{T}$ has the form
\begin{equation}
\lambda_{T}=\lambda-\frac{3}{16\pi^{2}\sigma^{4}}\lambda(m,T)
\end{equation}
and  we have defined $\lambda(m,T)$ as
\begin{equation*}
\lambda(m,T)=2m_{W}^{4}\ln\frac{m_{W}^{2}}{a_{b}T^{2}}+m_{Z}^{4}\ln\frac{m_{Z}^{2}}{a_{b}T^{2}}
-4m_{t}^{4}\ln\frac{m_{t}^{2}}{a_{f}T^{2}}~.
\end{equation*}               
In this expression for $\lambda_{T}$ we have introduced different constants
such that $\ln a_{f}=2\ln\pi-2\gamma\simeq 1.14$ and 
$\ln a_{b}=2\ln4\pi-2\gamma\simeq 3.91$, since in this way the term 
$-3/2B\phi^{4}$ of 
the zero temperature part cancels an equal term with positive sign which
appears in the finite temperature part.

In order to investigate the nature of the EWPT, the 
evolution of the effective potential with temperature will be the most 
important tool, but before to discuss how the phase transition proceeds, we 
need to check the
validity of the high temperature approximation. The evolution of the 
effective potential using the exact expressions and the high temperature 
approximation
for a range of temperatures is given in the next Fig.~\ref{Fig:51} 
The calculation of the integrals has been carried out by using again the same
numerical methods as in Section~\ref{Sect:Veff-HT}. 
For this calculation 
we have used the masses of the vector bosons as they are given in the 
review of the particle properties \cite{Hikasa:je}, so  we have taken
the mass of the $W$ boson as $m_{W}=91.173\,\textrm{GeV}$ and the $Z$ boson mass as 
$m_{Z}=80.22\,\textrm{GeV}$. For the specific case which we examine in this 
graph we have set the Higgs mass $m_{H}=50\,\textrm{GeV}$ and the top quark mass
$m_{t}=120\,\textrm{GeV}$.  The shape of the curves indicates a first order phase 
transition since there
are two local minima which become degenerate at a
temperature $T_{c}=85.935\,\textrm{GeV}$. The minimum of the effective potential at
the broken symmetry phase appears at a value of 
the field $\phi_{c}\approx 80\,\textrm{GeV}$.
As we can observe in Fig.~\ref{Fig:51} the high temperature expansion of the integrals
is in good agreement with the exact calculation. This result was expected
since as we have shown in the previous section, the high temperature
expansion is a good approximation to the exact calculation of the effective
potential.

\begin{figure}
\includegraphics[scale=0.45]{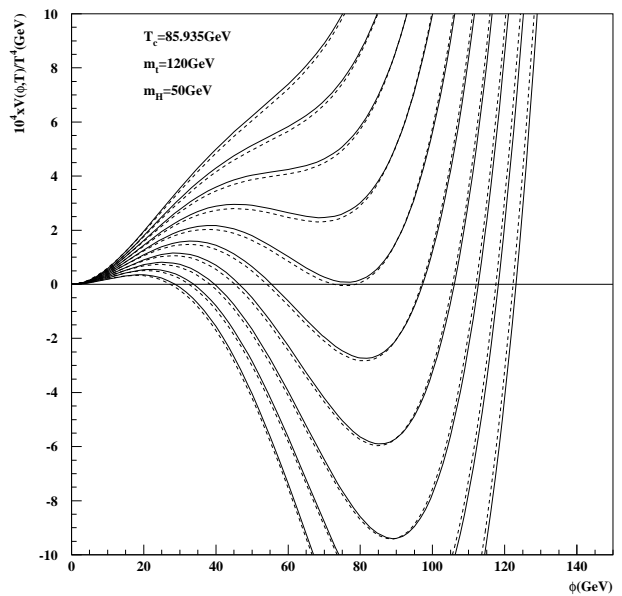}  
\caption{\label{Fig:51}The finite temperature effective potential of the standard model
for a range of temperatures 
$85.21\,\textrm{GeV}\leq T\leq 86.51\,\textrm{GeV}$ in steps of $0.145\,\textrm{GeV}$. The dashed line 
corresponds to the high temperature 
approximation, while the continuous line represents the evolution 
of the potential using the exact expressions. The Higgs mass is $m_{H}=50\,\textrm{GeV}$ 
and the top mass $m_{t}=120\,\textrm{GeV}$. The two local minima become degenerate 
at a temperature $T_{c}=85.935\,\textrm{GeV}$ and the value of the field at the second
minimum is  $\phi_{c}\approx 80\,\textrm{GeV}$.}
\end{figure}

\subsection{\label{Sect:SMPT} The Phase Transition}

\subsubsection{Evolution of the Potential}
  
In order to understand how the phase transition proceeds, we need to 
investigate the evolution of the effective potential with temperature.
As it has become clear from the analysis given in the previous section, we
can rely on the
high temperature approximation of the one loop effective potential 
in the regime of interest.

At very high temperatures the free energy density of the system, which is given
in Eq.~(\ref{Eq:v(phi,T)}), appears to 
have only one minimum, that at $\phi=0$. As the system cools down 
the effective potential acquires an extra minimum
at a value of the field 
\begin{equation}
\phi=\frac{3ET}{2\lambda_{T}}
\end{equation}  
which appears as an inflection point at a temperature 
\begin{equation}
T_{1}^{2}=\frac{T^{2}_{0}}{1-9E^{2}/8\lambda_{T_1}D}.
\end{equation}
As the temperature is lowered this minimum becomes degenerate with the other
one  at
$\phi=0$ and is separated from that with a potential barrier.  This happens
at a temperature $T_{c}$, when the quadratic equation  resulting from Eq.~(\ref{Eq:v(phi,T)})
dividing it by $\phi^{2}$ has two real equal roots. This is called
the critical temperature and is given by
\begin{equation}
T^{2}_{c}=\frac{T_{0}^{2}}{1-E^{2}/\lambda_{T_c}D}.
\end{equation}
When the system reaches the critical temperature, the value 
of the field at the broken symmetry phase is given by
\begin{equation}
\phi_{c}=\frac{2 E T_{c}}{\lambda_{T_{c}}}.
\end{equation}
The height of
the barrier describes the strength of the transition, if it is high then
the transition is strongly first order, whereas it is weak if the barrier
is small. 
In this case the phase transition can take place via barrier penetration or
tunnelling. If tunnelling rates are small there is supercooling until the
curvature of the potential becomes negative in the origin. This produces a
departure from thermal equilibrium 
and the phase transition proceeds by classical rolling of the $\phi=0$ vacuum
to $\phi=\sigma$ one. This  happens when the value of the field at the second
minimum becomes equal to
\begin{equation}
\phi_{0}=\frac{2 E T_{0}}{\lambda_{T_{0}}}~.
\end{equation}
In this case the phase transition is called spinodal.
  
\subsubsection{Unknown Parameters}
                                                        
\begin{figure}
\includegraphics[scale=0.45]{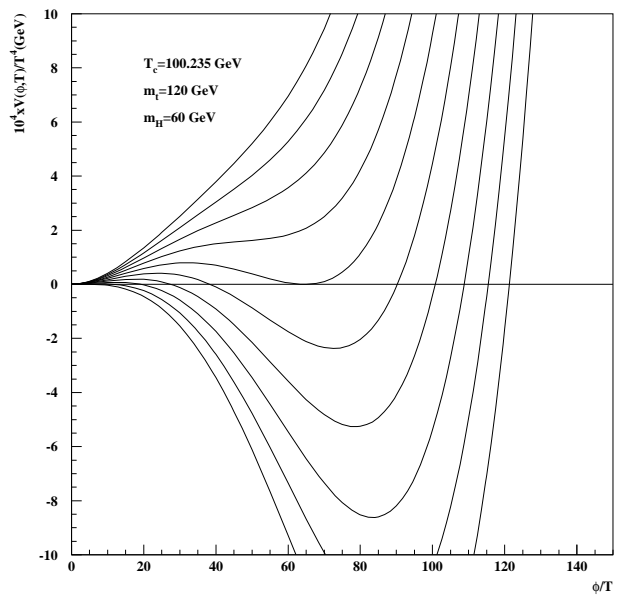} 
\caption{\label{Fig:52} The finite temperature effective potential at the high temperature
approximation for a Higgs mass 
$m_{H}=60\,\textrm{GeV}$
and top mass $m_{t}=120\,\textrm{GeV}$ for a range of temperatures 
$98.985\,\textrm{GeV}\leq T\leq 101.235\,\textrm{GeV}$ in steps of $0.25\,\textrm{GeV}$. The critical 
temperature is at 
$T_{c}=100.235\,\textrm{GeV}$ and the second minimum appears at a value of the field
$\phi_{c}\approx 65\,\textrm{GeV}$.}
\end{figure}

\begin{figure}
\includegraphics[scale=0.45]{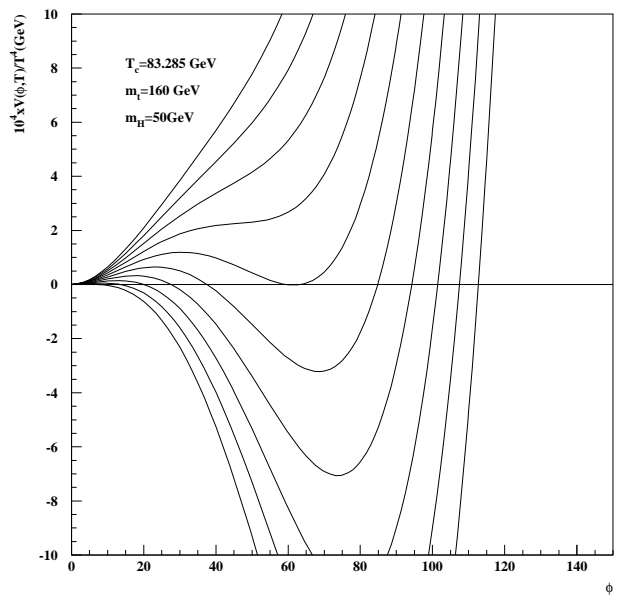} 
\caption{\label{Fig:53} The finite temperature effective potential at the high temperature
approximation for a Higgs mass 
$m_{H}=50\,\textrm{GeV}$
and top mass $m_{t}=160\,\textrm{GeV}$ for a range of temperatures 
$82.535\,\textrm{GeV}\leq T\leq 83.885\,\textrm{GeV}$ in steps of $0.15\,\textrm{GeV}$. The critical 
temperature is at 
$T_{c}=83.285\,\textrm{GeV}$ and the second minimum appears at a value of the field
$\phi_{c}\approx 60\,\textrm{GeV}$. }
\end{figure}

As we have stated already the transition is of the first order due to
the appearance of the term cubic in the field in the expression of the one 
loop effective potential. 
But if one
attempts to increase the Higgs boson or the top quark masses, the barrier 
between the two degenerate vacua appears smaller and at smaller values 
of the field $\phi$ than before. These arguments are illustrated 
for two particular cases in the next two 
figures: for $m_{H}=60\,\textrm{GeV}$, $m_{t}=120\,\textrm{GeV}$ in Fig.~\ref{Fig:52} and  
for $m_{H}=50\,\textrm{GeV}$, $m_{t}=160\,\textrm{GeV}$ in Fig.~\ref{Fig:53}~.

One can easily observe that the barrier in Fig.~\ref{Fig:52} appears lower than  
the one in Fig.~\ref{Fig:51}  and at a value of the field 
$\phi_{c}\approx 65\,\textrm{GeV}$. On the other hand  the two minima become
degenerate at a higher  critical temperature $T_{c}=100.235\,\textrm{GeV}$.         
If we continue to increase the Higgs mass, then 
things become confusing, the barrier appears too small and we cannot
reliably say that the phase transition is not of the second order. 

A similar situation appears if we increase the top mass, although the phase 
transition takes place at lower temperatures than in the case where we
increase the Higgs mass.  As  we can see in Fig.~\ref{Fig:53} the value of the 
field at the second minimum is $\phi_{c}\approx 60\,\textrm{GeV}$ but the critical
temperature appears much lower $T_{c}=83.285\,\textrm{GeV}$. 

\begin{figure}
\includegraphics[scale=0.45]{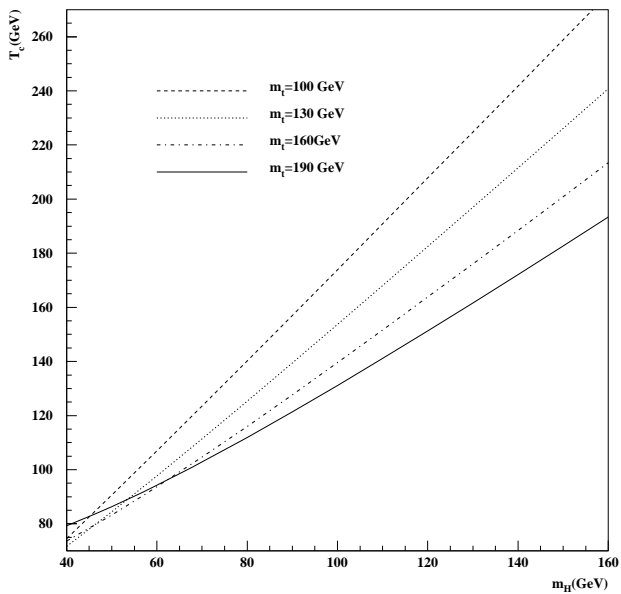} 
\caption{\label{Fig:54} The critical temperature of the electroweak phase transition as a
function of the Higgs mass for a range of top quark masses 
$100\,\textrm{GeV}\leq m_{t}\leq 190\,\textrm{GeV}$ in steps of $30\,\textrm{GeV}$. The dashed line
corresponds to $m_{t}=100\,\textrm{GeV}$ and the solid line to  
$m_{t}=190\,\textrm{GeV}$. We observe that as the top mass increases  the 
critical temperature decreases.}
\end{figure}

\begin{figure}
\includegraphics[scale=0.45]{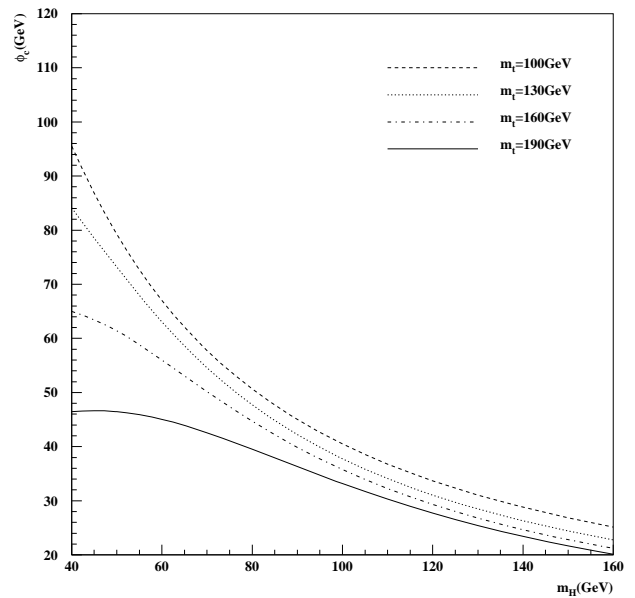} 
\caption{\label{Fig:55} The temperature dependent VEV of the Higgs field $\sigma(T)$ as a 
function of the Higgs mass for a range of top quark masses 
$100\,\textrm{GeV}\leq m_{t}\leq 190\,\textrm{GeV}$ in steps of $30\,\textrm{GeV}$.}
\end{figure}

In order to quantify the above arguments and illustrate  the dependence of 
the critical temperature to the particle masses,  the critical
temperature as a function of the Higgs mass for various top quark masses
in the range $100\,\textrm{GeV}\leq m_{t}\leq 190\,\textrm{GeV}$ is
given in Fig.~\ref{Fig:54}. As we can observe in this picture as  the top
mass increases  the critical temperature decreases.
The temperature dependent vacuum expectation (VEV) of the
field as a function of the Higgs mass for the same range of quark masses as 
before is given in Fig.~\ref{Fig:55}. A similar situation appears in this 
picture, as the top quark mass increases the VEV of the Higgs field decreases.
But as we can observe in Fig.~\ref{Fig:55}, the VEV  of the Higgs
field is not so sensitive to changes of the top mass if the Higgs mass is big
enough. We can see in Fig.~\ref{Fig:55}, that for Higgs masses larger than 
$100\,\textrm{GeV}$, the curves representing the various top masses are 
very close each other.

The unknown parameters of the theory cause a lot of uncertainty in a proper
study of the nature of the electroweak phase transition and
there  are a lot of contradictory claims about the order and the
strength  of the transition in the current literature. {\textit But for 
Higgs and top masses of an order of
magnitude close to the current experimental limits  which 
are about  $64\,\textrm{GeV}$ and $130\,\textrm{GeV}$ respectively, the one loop calculation 
of the effective potential 
suggests that the phase transition
is of the first order and takes place at a critical temperature 
about $100\,\textrm{GeV}$}.

\begin{figure}
\includegraphics[scale=0.45]{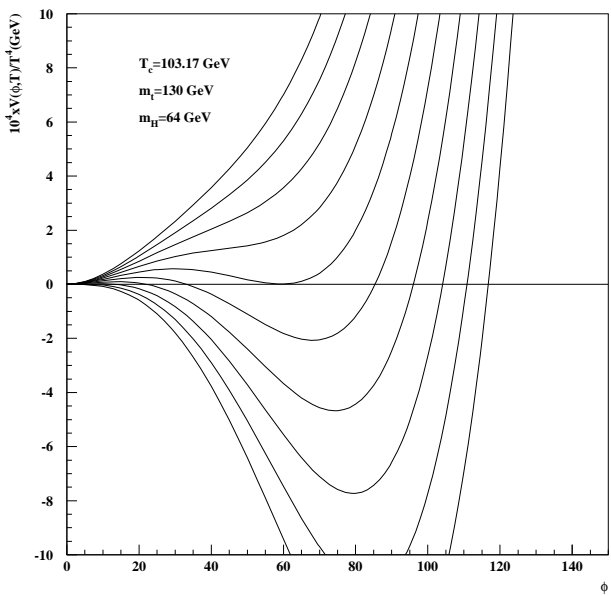} 
\caption{\label{Fig:56} Evolution of the finite temperature effective potential at the 
high temperature limit for a range of temperatures 
$101.92\,\textrm{GeV}\leq T\leq 104.17\,\textrm{GeV}$ in equal steps of $0.25\,\textrm{GeV}$. The Higgs 
mass is $m_{H}=64\; GeV$ and the top 
mass $m_{t}=130\; GeV$. The two minima become degenerate at a critical
temperature  $T_{c}=103.17\; GeV$ and at a value of the field about 
$60\,\textrm{GeV}$.}
\end{figure}

In order to illustrate this argument in  Fig.~\ref{Fig:56} we give the 
evolution of the effective potential  with temperature where in
this particular case  we have used the lower experimental 
limits for the
particle masses  $m_{H}=64\,\textrm{GeV}$ and $m_{t}=130\,\textrm{GeV}$ 
\cite{Abachi:1994je, Luth:1993ey}. The two minima 
become degenerate at a temperature 
$T_{c}=103.17\,\textrm{GeV}$ and the value of the field at the second minimum appears
at $\phi_{c}\approx 60 \,\textrm{GeV}.$

The phase transition is of the first order and  in this way
the departure of thermal equilibrium is satisfied but the vacuum expectation 
value of the scalar field  after the completion of the phase transition is 
small and becomes even smaller if the particles are heavier. This imposes 
constraints for baryogenesis to happen at the electroweak phase 
transition, since as it was stated in the third section, in order   to prevent 
the erasure due to sphaleron processes of the baryons generated during the
transition, the condition $\sigma(T_{c})/T_{c}\geq 1$ must be  satisfied. 
But this condition is clearly not satisfied  for Higgs and top 
masses above the current experimental limit.

\begin{figure}
\includegraphics[scale=0.45]{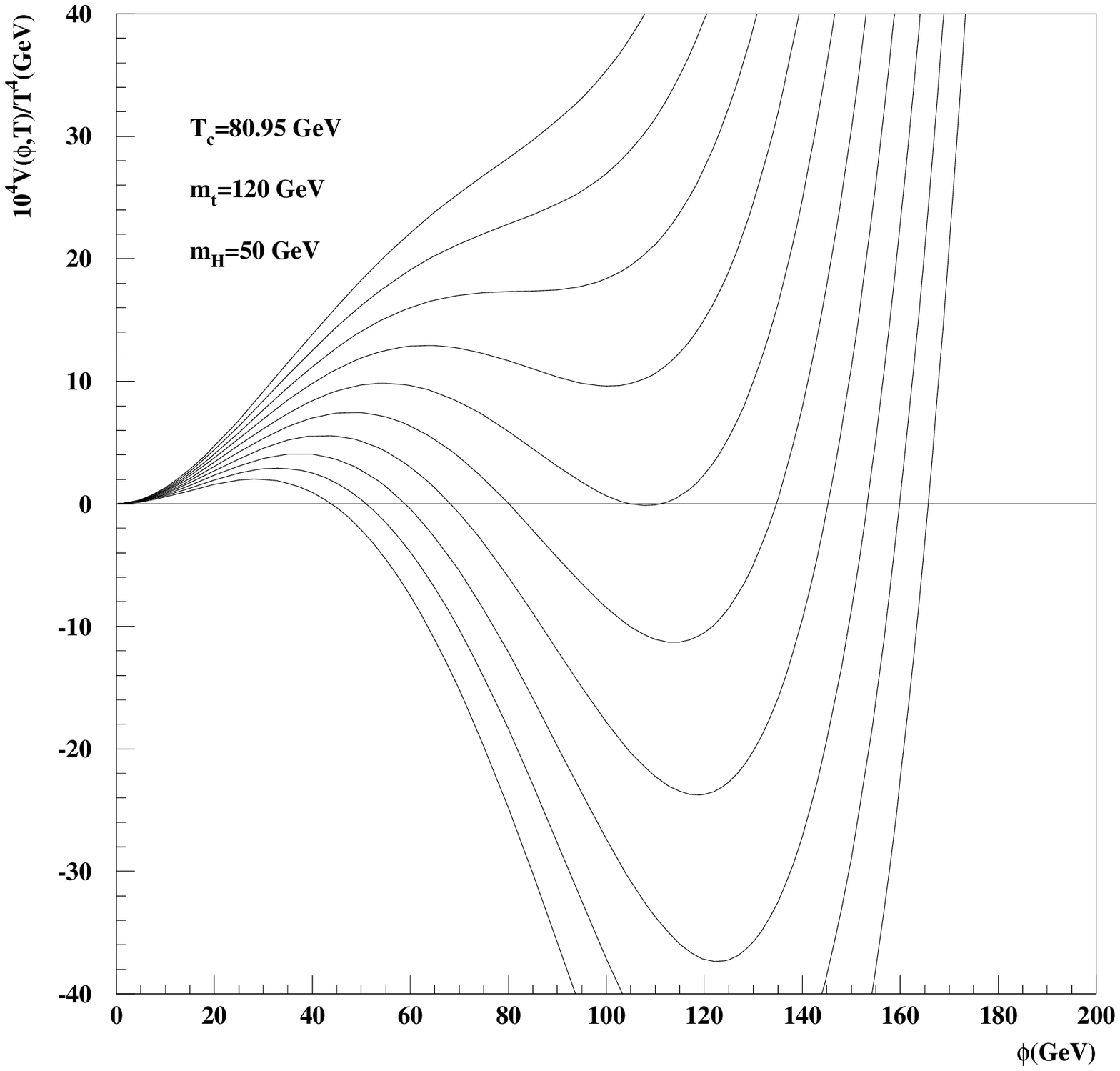} 
\caption{\label{Fig:57} The finite temperature effective potential ignoring zero temperature
quantum corrections, for a range of temperatures
$79.7\,\textrm{GeV}\leq T\leq81.95\,\textrm{GeV}$ in steps of $0.25\,\textrm{GeV}$. The two minima 
become degenerate at a critical
temperature $T_{c}=80.95\,\textrm{GeV}$ and at a value of the field $\phi_{c}\approx
110\,\textrm{GeV}$. We have taken the Higgs mass $m_{H}=50\,\textrm{GeV}$ and the top mass
$m_{t}=120\,\textrm{GeV}$.}
\end{figure}

A very different situation appears if one ignores the zero temperature 
quantum  effects. In this case the effective potential  is the sum of the 
classical potential plus the fermion and boson integral contributions which
we calculate numerically. The height  of the barrier is much bigger now, so 
the phase transition appears more strongly first order than before
and takes place at a lower critical temperature. 
Our investigation confirms 
similar results which are presented in a recent work by  Vinh Mau 
\cite{VinhMau:cy}.
A particular case where we have set the Higgs mass $m_{H}=50\,\textrm{GeV}$ and the top
mass $m_{t}=120\,\textrm{GeV}$ is given in Fig.~\ref{Fig:57}  where  the two 
degenerate minima
appear at a temperature $T_{c}=80.95\,\textrm{GeV}$. Although the graphs 
in Fig.~\ref{Fig:51} and Fig.~\ref{Fig:57}
are given in different scales, one can easily observe that the barrier in the
second case, where we have ignored the zero temperature 
quantum corrections, appears nearly four times higher than in the 
first case where we include them. 
\begin{figure}
\includegraphics[scale=0.45]{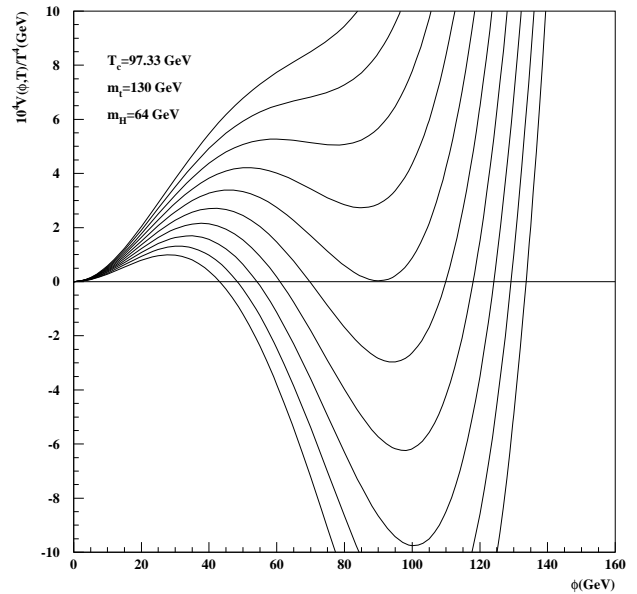} 
\caption{\label{Fig:58} The finite temperature effective potential ignoring zero temperature
quantum corrections, for a range of temperatures
$96.58\,\textrm{GeV}\leq T\leq 97.83\,\textrm{GeV}$ in steps of $0.15\,\textrm{GeV}$. The two minima 
become degenerate at a critical
temperature $T_{c}=97.33\,\textrm{GeV}$ and at a value of the field $\phi_{c}\approx
90\,\textrm{GeV}$. We have taken the Higgs mass $m_{H}=64\,\textrm{GeV}$ and the top mass
$m_{t}=130\,\textrm{GeV}$.}
\end{figure}
But if we increase the Higgs mass a similar situation appears as before when 
we had included the zero temperature part.  In the particular case  given in
Fig.~\ref{Fig:58}, where we use the current experimental limits for the particle
masses, it is obvious that the barrier is smaller than the one in
Fig.~\ref{Fig:57}, while on the other hand the critical temperature is much higher. 

If we compare the evolution of the potential given in Fig.~\ref{Fig:58} with
the case given in Fig.~\ref{Fig:56} we can easily observe that in  
the last case where we have ignored the zero temperature quantum 
corrections the phase transition appears more strongly first order.
Although the vacuum expectation value of the field is much bigger now
$\phi_{c}\approx 90\,\textrm{GeV}$, the critical temperature is 
$T_{c}=97.33\,\textrm{GeV}$ and thus the condition
$\sigma(T_{c})/T_{c} \geq 1 $ is not satisfied.

\subsection{\label{Sect:Bubble-Nucleation}Bubble Nucleation}   

As in the case of the boiling  water,
first order phase transitions  proceed via formation and subsequent expansion
of bubbles of the new phase within the  old one.
As the temperature is lowered  below the critical temperature $T_{c}$, 
the electroweak phase
transition is anticipated to proceed by formation of bubbles of the stable
phase (broken symmetry phase $\phi\neq 0$) within the metastable one (symmetric
phase $\phi=0$). 
The rate of nucleation of the broken symmetry phase is increasingly 
important, but a complete analysis of the dynamics of first order phase
transitions taking into account the expansion of the universe is still
lacking.
The formation and propagation of the bubbles of the new phase
has received much attention in the current bibliography and there have been a
lot of investigations since the early works 
by Linde \cite{Linde:tt} and Langer \cite{Langer:bc}.

Determination of the baryon asymmetry, which is produced during the
phase transition, requires  knowledge not only how the bubbles are produced, but
also how  they evolve in time until they coalesce and fill the universe with 
the new broken symmetry phase. There are many problems  which are associated 
with the propagation of the new phase bubbles  as for example  the wall 
velocity and the wall size.  
In many models the baryon asymmetry in produced in the bubble wall, the region
which interpolates between the stable and the unstable phase of matter. 
There are two important limiting cases in this problem, the thin 
wall and the thick wall approximations \cite{Anderson:1991zb,Dine:1992wr}.
In the thin wall approximation the difference between the two minima of the
effective potential is much smaller than the height of the barrier between
them. The radius of the bubble  at the moment of formation is much larger than 
the size of the wall and the phase transition proceeds with  small 
supercooling. In the thick wall case the difference in depth between the two 
minima is much bigger and a large amount of supercooling is needed for the
phase transition to proceed. 

In a recent investigation of bubble
nucleation rates in first order phase transitions, it was shown that the model
provides a sound quantitative framework for the determination of
supercooling in the electroweak phase transition \cite{Cottingham:rv}. The 
evolution in time of the broken symmetry phase bubbles is a live current 
research topic. Preliminary
investigations of this evolution in the framework of the Langevin equation
seem to indicate that the transition is completed well before the universe
reaches the spinodal point, but still a large amount of supercooling 
is needed for this to happen \cite{Kalafatis:1994,Cottingham:1994ax}.

%% file: chap6.tex
\section{\label{Sect:Summary}Summary}

\subsection{\label{Sect:Discussion}Discussion}

Throughout this work it is apparent that the standard electroweak
model provides the framework where all the Sakharov conditions 
for a possible explanation of the observed baryon excess could be satisfied. 
However, the question how large an asymmetry is actually generated still holds,
and it seems unlikely that an asymmetry of the experimentally observed order of magnitude
can be obtained within the framework of the minimal standard model with one
Higgs doublet. The main reasons that contradict the minimal model come from the 
magnitude of $CP$ violation and the experimental limits on Higgs mass.

The effects of $CP$ violation in the minimal standard model coming from the 
CKM phase are much too small to explain the observed asymmetry
\cite{Shaposhnikov:1991pd} and the possible enhancement suggested by the same
author \cite{Shaposhnikov:tw,Shaposhnikov:1987pf}, seem to be not
acceptable in general \cite{Dine:1991ck,Cohen:1993nk}.     
Incorporating additional $CP$ violation into the standard model requires 
additional matter fields and there are a number of extensions of the 
minimal model one can consider \cite{Cohen:1991iu}. Amongst
them the model with two Higgs doublets has received much attention recently.

Baryon number violation is satisfied  in the minimal model and the rate of baryon
violating processes at high temperatures is large enough to produce the 
baryonic asymmetry \cite{Kuzmin:1985mm}. However any baryons
produced during the EWPT should survive until the present. This requires
a large VEV for the Higgs field after the EWPT is completed. This restriction
comes from the requirement that the sphaleron mass in the broken 
symmetry phase must be large enough to strongly suppress a washing out of the
baryons just created at the transition. This sphaleron constraint places
a theoretical upper bound on Higgs mass, which has been estimated to be about
$45\,\textrm{GeV}$ \cite{Shaposhnikov:tw, Shaposhnikov:1987pf} but this is not 
supported by the latest experimental data, where the lowest bound on the 
Higgs mass is about $64\,\textrm{GeV}$ \cite{Luth:1993ey}.

One possible way to increase the theoretical upper bound on the Higgs mass
is to extend the scalar sector of the theory as for   
example in the work by Anderson and Hall \cite{Anderson:1991zb}
where a simple scalar is introduced and the author obtained an upper bound
of about $50\,\textrm{GeV}$ for the Higgs mass or to consider more complicated 
cases as it is the two doublets model. In a recent paper 
\cite{Choi:cv}, it was shown that addition of a real Higgs singlet 
results in a 
strongly first order phase transition and they obtained an upper
bound for the Higgs about $60\,\textrm{GeV}$. But it is easy to see that the current
experimental data invalidate the results in both  cases. Most promising
seem to be the two doublets model and  has been suggested  
by  many authors since, since it provides the possibility
to enhance the rate of $CP$ violation as well through a  $CP$ violating
relative phase between the two doublets \cite{Turok:1992ar}. An analysis of the 
phase transition
in the two doublets case is given by Turok and Zadrozny 
\cite{Turok:1990in} where the authors obtained an upper limit for 
the lightest Higgs boson of about $120\,\textrm{GeV}$.

The condition that the universe must be out of thermal equilibrium is satisfied
if the EWPT is of the first order and we have shown that 
the one loop calculation of the effective potential predicts a first order 
phase transition.  This conclusion remains valid 
when one includes higher order graphs, as  for example
the ``ring graphs'', which are introduced in order to cure infrared problems 
as has been discussed in recent papers by Carrington \cite{Carrington:1991hz}
and Espinosa \textit{et al.} \cite{Espinosa:1992gq}. But there are some  
questions about the strength of the
transition with some authors to claim that the transition is strongly
first order as for example Shaposhnikov \cite{Shaposhnikov:1991cu} who obtained
a linear term in the expression of the effective potential and others who 
claim that it takes place weakly. 
The appearance of a linear term into the effective potential but with 
opposite sign is also reported by Brahm and Hsu \cite{Brahm:1991nh} and as
a result the phase transition ceases to be first order for some 
temperatures.  Another point of view is discussed by 
Dine \textit{et al.} \cite{Dine:1992wr}, where the authors suggested that no
linear terms appear into the effective potential, since on the other hand
higher order corrections lead to a significant modification of the one loop
result and the phase transition is weakly first order. 
They have obtained a cubic term  into the final expression 
of the effective potential at the high temperature limit, which is diminished
by a factor of 2/3 and this would weaken the phase transition.
A different possibility is given in \cite{Choi:cv}, where according to the 
authors the strength of the transition due 
to the appearance of tree level cubic terms in the Higgs potential, in
contrast to the one loop effective potential where the cubic terms appear
at finite temperature only. They estimated that the transition is strong 
enough to prevent
the erasure due to sphaleron of the baryon asymmetry but the bound they
obtained on Higgs mass contradicts the current data.

\subsection{\label{Sect:Conclusions}Concluding Remarks}

In order to understand the nature of the electroweak  phase transition at the
one loop  level,  in this work we have examined the effective
potential in the minimal standard model with one Higgs doublet.
In our investigation  we have shown that at this level,  the high 
temperature 
expansion of the integrals at the finite temperature effective potential 
is in good agreement with the exact calculation and that the electroweak 
phase transition is of the first order. The first order character of the 
transition is due to the appearance of a term cubic in the field into 
the expression of the effective potential at the high temperature limit.

But even at this simple level things seem to be too complicated. The unknown 
parameters of the theory, the Higgs and the top quark masses, cause a lot of
uncertainty in the calculation of the effective potential as became evident 
from our analysis of the behaviour of the effective potential for different
Higgs and top masses. As a result the strength of the transition cannot be
described properly, but we can say safely that it is of the first order 
as long as
the cubic term in the field appears into the expression of the finite
temperature effective potential.

All the above considerations lead us to conclude that there is a lot
of work to be done, before the minimal standard electroweak model or its 
extensions can be established  as the model providing the solution to the
baryogenesis problem and satisfy the Sakharov conditions. 
Only if the two missing ingredients of the standard electroweak model, the 
Higgs, one or more, and the top quark will be detected  and their masses 
will be measured properly, there will be a clarification of these 
questions. On the other hand the reliability of the  loop expansion of
the effective potential has to be more seriously examined in the future and
as it is stressed in \cite{Dine:1992wr} \textit{``without a proper study of the higher
order corrections to the effective potential, one may be unable to make 
any conclusions concerning the possibility of baryogenesis in the standard
model''}.

%% file: remarks.tex
\subsection{\label{Sect:Comments}Some comments about recent progress}

Since the time when the work presented in the previous sections was initially written, there has 
been been quite some progress on electroweak baryogenesis and a substantial number of investigations  
have been published on the subject. A search in 
SLAC--SPIRES for \textit{title electroweak baryogenesis and date after 1994} finds more than 120 
research papers. Recent progress on the subject has been reviewed by Trodden \cite{Trodden:1998ym}
and also by Riotto and Trodden \cite{Riotto:1999yt}, while a latest account 
is given by Dine and Kusenko \cite{Dine:2003ax}. In this section we 
are trying to update some of the information given in the previous sections.

In Section~\ref{Sect:Baryon-Asymmetry} we have given an estimation of the baryon 
asymmetry based on information published
at the period when this work was initially written. However, according to recent measurments 
of the fluctuations of the cosmic microwave background 
by the WMAP collaboration \cite{Bennett:2003bz}, the baryon 
asymmetry is known to a $5\%$ accuracy as 
\begin{equation}
B=\frac{n_{B}}{s} \sim(6.1 ^{+0.3}_{-0.2})\times 10^{-10}~.
\end{equation}

I have tried to present in some detail in Section~\ref{Sect:Veff}, how we can calculate 
the effective potential at zero and finite temperature reffering to earlier papers or 
books. There is however a recent paper by Quiros \cite{Quiros:1999jp}
where many different aspects of the effective
potential are discussed in detail. 

We have concluded in Section~\ref{Sect:Conclusions}, that the strength of the phase transition depends
on the unknown parameters of the model which are the top and the Higgs masses. However, the top 
mass is not longer unknown. According to latest data of 
the Particle Data Group \cite{Hagiwara:fs}, the top mass is 
evaluated using the avegare of five different measurements as $m_t=174.3 \pm 5.1\,\textrm{GeV}$. The Higgs 
mass is still a puzzle and the 
lower current experimental limit (assuming a SM Higgs) is  
about $m_H=115\,\textrm GeV$ \cite{Hagiwara:fs}. 

Also in Section~\ref{Sect:Conclusions} it is mentioned that a proper study of higher order
loops in the calculation of the effective potential is needed. While writing those lines
I was not aware that there had been already published contemporary investigations on the effect 
of higher loops. Amelino--Camelia \cite{Amelino-Camelia:1994kd}, based 
on the Cornwall--Jackiw--Tomboulis method
of composite operators \cite{Cornwall:1974vz}, is discussing a self--consistent improvement 
of the effective poential where the daisy and superdaisy graphs are taken into account. Also
Arnold and Espinosa \cite{Arnold:1992rz}
have calculated the effective potential beyond the leading order. In both cases, the first 
order nature of the phase transition persists, however the strength is also questionable.

We have concluded that the minimal model at our one loop calculation does not really offers
a satisfactory solution for baryogenesis to happen during the electroweak
phase transition. As it was already mentioned 
in Section~\ref{Sect:Discussion}, models which go beyond the SM seemed to be more promising. This 
conclusion is validated in the recent review by 
Dine and Kusenko \cite{Dine:2003ax}, where various models are explored. 
As it is discussed in this work, eventually the
minimal SM baryogenesis should be ruled out. The main reason for this conclusion is, 
that according the results of a number of investigations based on numerical 
simulations \cite{Kajantie:1996mn,Kajantie:1996qd,Kajantie:1998rz,Rummukainen:1998as}, the 
electroweak phase transitions is not first order anymore, but 
it turns to a smooth crossover. However, SUSY extensions of the SM are still viable even if
these models are still questionable. The sphaleron bound imposes  very strict conditions in 
order for  baryogenesis to happen during  the 
the electroweak phase transition. This is because of the Higgs mass 
which is unknown. Experimental searches have failed to establish the SM Higgs mass or
the mass of its minimal supersymmetric partners. The lower limits on the Higgs mass
have made models like the MSSM to run into dufficulties. However, in a recent analysis 
by Servant \cite{Servant:2001jh}, it is shown 
that there is still a possibility for a way out 
if one alters  the Friedmann equations. 

The whole idea of electroweak baryogenesis is still very attractive mainly 
because of the restrictions which are imposed by inflation. As it was stated 
already at the beginning of 
this work, baryogenesis has to happen after inflation in order to
prevent the washout of any baryon asymmetry generated previously. We have referred 
to inflation only briefly at the introduction. Inflation
is on its own an enormous subject, which there is no way that we would be able to discuss it 
here properly. The recent status of inflation models is reviewed by 
Lyth and Riotto \cite{Lyth:1998xn} and also by Kolb \cite{Kolb:1999ar}
and Brandenberger \cite{Brandenberger:1999sw}. The possibility of electroweak baryogenesis
with new hybrid inflation models is investigated in a recent work by
Copeland, Lyth, Rajantie and Trodden \cite{Copeland:2001qw}. They propose novel mechanisms 
where the baryon asymmetry is generated after inflation.

In order for the three Sakharov conditions to be satisfied, we have demanded 
that the electroweak phase transition has to be of the first order. In such a case,
a mechanism for the transition to proceed is by bubble nucleation. We have  already 
discussed briefly this mechanism in 
Section~\ref{Sect:Bubble-Nucleation}. Recent progress on the dynamics of 
bubble nucleation is discussed by Bergner 
and Bettencourt \cite{Bergner:2002we}, Moore and Rummukainen \cite{Moore:2000jw} and others
\cite{Surig:1997ne,Baacke:1995bw}.

For simplicity in the calculations in this work (and many others), the chemical potential
has been set equal to zero. There is however a 
very recent work on electroweak baryogenesis by Gynther \cite{Gynther:2003za}, where the relation between
chemical potential and the Higgs mass is investigated.

\subsection{A final note about this work}

This work is different in many respects from the original report which has been submitted
in a form of a thesis in order to fulfil the requirements of a MSc degree 
in theoretical physics at 
the University of Manchester on April 1994. However, the essential physics results are
basically the same. I decided to post this in the arXiv,
even if it is quite out of date, hoping that it might be useful to someone. I have not altered the
basic structure of the text, only tried to correct some typos, updated a bit some older 
latex forms and in order to save pages I have used revtex4 format. This means that I had also to 
alter some equations in order to fit 
in the two column style. I had also to reproduce some of the figures. However, the bibliography file 
has been produced again
using the format of the SLAC--SPIRES, and this is because I think that in SPIRES
they do an excellent job and deserve some help with the citations. I have 
also tried  to include some references regarding recent works on the subject. However, since I do not 
really work on baryogenesis now, and therefore I do not consider myself as an expert 
on it, this attempt definitely cannot be complete. Therefore, I apologize in advance if I have omitted 
important work, and thank all these people that I am referred to, for the things which I have learned 
going through their work.

%% file: acknow.tex
\begin{acknowledgments}

These are not the original acknowledgments written for
the thesis report, where I had to thank all the people who supported me 
during that period, and I do thank them
again, but the list is too long to be repeated here. However, 
I would like, once again, to express my graditude to Mike Birse who was my advisor for this work,
for his constant support encouraging and patience for my sometimes slow progress. Dimitri Kalafatis who
now has quited physics for a job on Risk Analysis, had been a great
and costant source of support and suggestions. And I should not also
forget my teachers at the time, Graham Shaw, Pat Buttle, Tony Phillips 
and John Storrow who introduced me to 
the ideas of quantum field theory and gauge theories. Unfortunately, Tony and John are no longer
with us, so I feel that I should dedicate this work to their memory. Finally, I would like to 
thank Eef van Beveren
who offered me a job on his project here in Coimbra, and  has allowed me to use some of my time to update 
this work. I also would like to thank Eef and Alex Blin for the numerous discussions on physics and not
only that we had during the last year. Especially Alex has made couple of comments about 
missleading points in this work as well. I hope that eventually I have managed to 
clarify properly. Financial support of the 
\textit{Funda\c{c}\~{a}o para a Ci\^{e}ncia e a Tecnologia}
of the \textit{Minist\'{e}rio da Ci\^{e}ncia e da Tecnologia} of Portugal, under the 
contract CERN/FIS/43697/2001 during my 
stay in Portugal is also acknowledged.

\end{acknowledgments}

%% file: NP-MScThesis.bbl
\begin{thebibliography}{99}
\bibitem{top} The mass of the top is no longer unknown, but according to 
latest data from  PDG \cite{Hagiwara:fs}
$m_t=174.3 \pm 5.1 \;\textrm{GeV}$.
\end{thebibliography}
